\documentclass[a4paper,reqno,superscriptaddress,nofootinbib, pra, aps,11pt]{revtex4-2}
\usepackage[centertags]{amsmath}
\usepackage{amsfonts}
\usepackage{amssymb}
\usepackage{amsthm}
\usepackage{titlesec}
\usepackage{newlfont}
\usepackage{stmaryrd}
\usepackage{mathrsfs}
\usepackage{mathtools}
\usepackage{graphicx}
\usepackage{enumerate}
\usepackage{todonotes}
\usepackage{color}
\usepackage{float}
\usepackage{orcidlink}
\usepackage{caption}
\usepackage{subcaption}
\usepackage{caption}
\usepackage{euscript}
\usepackage[margin=1in]{geometry}

\usepackage{tikz}
\usepackage{pgf}
\usetikzlibrary{positioning,fit,calc}
\usetikzlibrary{arrows,automata}
\usepackage{wrapfig}
\usepackage{amscd}
\usepackage{cancel}
\usepackage{hhline}
\usepackage{import}
 
\usepackage[commandnameprefix=always]{changes}
\usepackage{changes}
\date{\today}

\theoremstyle{plain}

\theoremstyle{definition}

\theoremstyle{remark}

\newcommand{\fig}{Fig.\;}

\newcommand{\commentA}[1]{[$\mathbb{A}$:{\ {\bf #1}}]}

\let\al=\alpha   
\let\ve=\varepsilon   
 \let\la=\lambda

\let\oldsqrt\sqrt
\def\sqrt{\mathpalette\DHLhksqrt}
\def\DHLhksqrt#1#2{
	\setbox0=\hbox{$#1\oldsqrt{#2\,}$}\dimen0=\ht0
	\advance\dimen0-0.2\ht0
	\setbox2=\hbox{\vrule height\ht0 depth -\dimen0}
	{\box0\lower0.4pt\box2}}

\newcommand{\id}{\textrm{d}}

\let\ve=\varepsilon

\usepackage{url}
\usepackage{hyperref}
\usepackage{lipsum}
\usepackage{mathtools}
\usepackage{lmodern}
\usepackage{anyfontsize}

\usepackage{tikz}
\usepackage{bigints}

 \DeclareFontFamily{U}{wncy}{}
    \DeclareFontShape{U}{wncy}{m}{n}{<->wncyr10}{}
    \DeclareSymbolFont{mcy}{U}{wncy}{m}{n}
    \DeclareMathSymbol{\Sh}{\mathord}{mcy}{"58}

\usepackage{tikz}
\usepackage{pgfplots}
\pgfplotsset{compat =1.18, width = 16 cm, height = 9 cm}
\usetikzlibrary{3d,decorations.markings}
\usepgfplotslibrary{fillbetween}

\usepackage{amsmath, amssymb}
\usepackage{tikz}
\usetikzlibrary{decorations.markings}
\usepackage{xcolor}
\usetikzlibrary{decorations.pathmorphing, arrows.meta}
\usepackage[svgnames]{xcolor}
\usepackage{enumitem}   

\usepackage{xcolor,soul}

\usepackage{CJKutf8}
\usepackage{tikz}
\usetikzlibrary{positioning}
\usepackage{amsthm}

\usepackage{amsthm}

\newtheoremstyle{example} 
  {\topsep}{\topsep}
  {\upshape}
  {}
  {\slshape \scshape}
  {}
  {\newline}
  {\thmname{#1}\ \thmnumber{#2}: \thmnote{#3}} 

\theoremstyle{example}
\newtheorem{example}{Example}

\newtheoremstyle{example_contd}
  {\topsep}{\topsep}
  {\upshape}
  {}
  {\slshape\scshape}
  {}
  {\newline}
  {\enspace Continuation of \thmname{#1}\ \thmnumber{#2}: \thmnote{#3}} 

\theoremstyle{example_contd}
\newtheorem{example_contd}{Example}

\begin{document}

\title{Quasistatic response for nonequilibrium processes: \hspace{10 cm} evaluating the Berry potential and curvature}

\author{Aaron Beyen \orcidlink{0000-0002-4341-7661}}
\author{Faezeh Khodabandehlou \orcidlink{https://orcid.org/0000-0001-8114-6105}}
\author{Christian Maes \orcidlink{0000-0002-0188-697X}}
\affiliation{Department of Physics and Astronomy, KU Leuven}
\date{}

\makeatletter
\def\@date{}
\makeatother

\begin{abstract}
We investigate how introducing slow, time-dependent perturbations to a steady, nonequilibrium process alters the expected (excess) values of important observables, such as the dynamical activity and entropy flux. When we make a cyclic thermodynamic transformation, the excesses are described in terms of a (geometric) Berry phase with corresponding Berry potential and Berry curvature quantifying the response. Focussing on Markov jump processes, we show how a non-zero Berry curvature leads to a breakdown of the thermodynamic Maxwell relations and of the Clausius heat theorem. We also present a variant of the Aharonov–Bohm effect in which the parameters follow a curve with vanishing Berry curvature, but the system still experiences a nonzero Berry phase. 
Finally, we identify (sufficient) no-localization conditions in terms of mean first-passage times under which the corresponding Berry potentials and curvatures vanish at absolute zero, extending, for arbitrary driving, \textit{e.g.} the case of vanishing heat capacity as for the Nernst postulate.
\end{abstract}
\maketitle


\section{Introduction}
Thermodynamic transformations 
are associated with quasistatic changes of control parameters of a system, such as coupling coefficients or environmental pressure and temperature. The idea is that these parameters can be varied so slowly compared to the relaxation time that the system remains in thermodynamic equilibrium for their instantaneous values \cite{Landaustatmech, pippard1966elements}. The same slow transformations apply to steady nonequilibrium systems where there is a constant driving or agitating force. The system now remains close to the instantaneous steady condition, and additional control parameters enter, such as the strength of a rotational force or of the agitation.  However, the effect of quasistatic changes can be quite different for steadily driven systems, and the response might not be expressible as the derivative of an appropriate free energy, \cite{Maes_2019}. In fact, the notion of {\it excess} enters when considering transformations between steady conditions, \cite{oono,Hatano_2001,epl,Komatsu_2008,MaesNetocny2015,jchemphys}. We have, for instance, directed and undirected currents of a certain type (particle, energy,...) with stationary values depending on the strength of external fields and temperature(s). When these parameters change to new values, the very change induces extra currents, and those are called excess currents. As a result, the quasistatic response around nonequilibrium steady states becomes more complicated but also richer. \\
In this paper, we investigate the excesses in entropy flux, heat, currents, and dynamical activity of Markov jump processes (verifying local detailed balance \cite{ldb} but not excluding strong driving or agitation). We obtain {\it geometric} expressions, containing the (equivalent of what has been called the) Berry phase, Berry potential and Berry curvature,  \cite{Berry1984, Berry_1985, wilczek1989geometric, hannay, Kariyado_2016, harrythirty, Sinitsyn_2007}. This identification with Berry’s formulation extends and unifies previous work on geometric phases in stochastic and dissipative systems, \cite{Chernyak_2009, stochastic1, Stochastic2, dissipativeattractors, dissipativeattractors2}. In particular, we show how the Berry curvature, as used here, measures the violation of (standard equilibrium thermodynamic) Maxwell relations \cite{Landaustatmech, pippard1966elements, Maes_2019}, and of the Clausius heat theorem.  We also construct an elementary analogue of the Aharonov-Bohm effect \cite{Abeffect, Sakurai_Napolitano_2020}, where a particle current exists in regions with no Berry curvature. We end with a discussion of low-temperature asymptotics where an extended Nernst postulate is obtained: Berry potentials vanish at absolute zero in the absence of localization.  The latter is specified, for all excesses, in terms of the relative boundedness of the same mean first-passage times.

\section{Excesses and their response}\label{first}
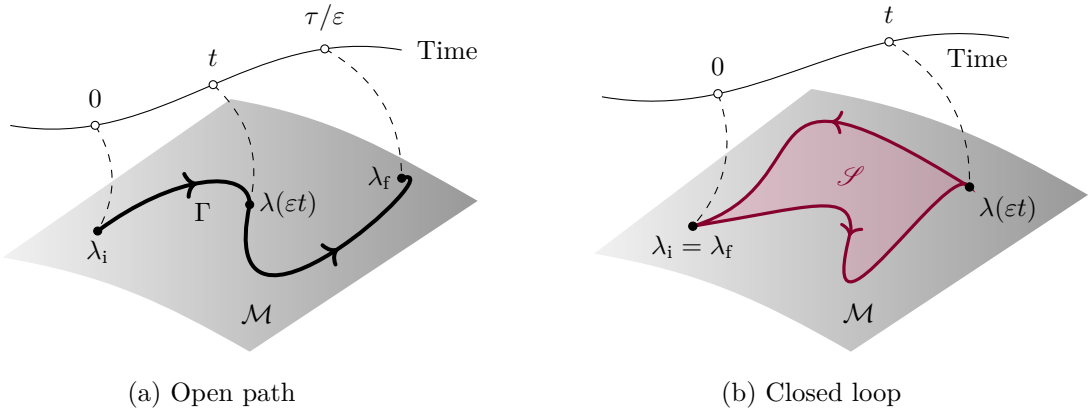
\begin{figure}[ht]  
\centering 
  \begin{subfigure}[t]{0.4\linewidth}
       \begin{tikzpicture}[scale = 0.5]
       \useasboundingbox (-1,-3) rectangle (11,4.5);
\path[name path=border1] (0,0) to[out=-10,in=150] (6,-2);
\path[name path=border2] (12,1) to[out=150,in=-10] (5.5,3.2);
\path[name path=redline] (0,-0.4) -- (12,1.5);
\path[name intersections={of=border1 and redline,by={a}}];
\path[name intersections={of=border2 and redline,by={b}}];

\shade[left color=gray!10,right color=gray!80] 
  (-0.5,-0.5) to[out=-10,in=150] (6,-3) -- (12,1) to[out=150,in=-10] (5.5,3.7) -- cycle;
\draw (-0.3,3) to[out=-10,in=170] 
  coordinate[pos=0.2] (aux1) 
  coordinate[pos=0.52] (aux2) 
  coordinate[pos=0.82] (aux3) (10,5);
\coordinate (lambda_i) at (2,0.2);
\coordinate (lambda_f) at (10,1.6);
\coordinate (midpoint) at (6,0.9);

\node[below] at (lambda_i) {$\lambda_{\mathrm{i}}$};
\node[left] at (lambda_f) {$\lambda_{\mathrm{f}}$};
\node[right] at (midpoint) {$\lambda(\ve t)$};
\node[below right] at (4.3,1.2) {$\Gamma$};

\tikzset{
    multiarrow/.style={
        decoration={markings,
            mark=at position 0.25 with {\arrow{>}},
            mark=at position 0.75 with {\arrow{>}}
        },
        postaction={decorate}
    }
}

\draw[line width=1.5pt, smooth, multiarrow]
   (lambda_i) .. controls +(2,1.5) and +(0,1.0) .. (midpoint)
             .. controls +(-1,-4.8) and +(1.5,0.5) .. (lambda_f);

\draw[dashed] (aux1)to[bend left]  (lambda_i);
\draw[dashed] (aux2)to[bend left]  (midpoint);
\draw[dashed] (aux3)to[bend left]  (lambda_f);

 \draw[fill=black] (lambda_i) circle (3pt);
 \draw[fill=black] (lambda_f) circle (3pt);
 \draw[fill=black] (midpoint) circle (3pt);


\foreach \coor in {1,2,3}
{
  \draw[fill=white] (aux\coor) circle (3pt);
}
\node[label=above:$0$] at (aux1) {};
\node[label=above:$t$] at (aux2) {};
\node[label=above:$\tau/\ve$] at (aux3) {};


\node[rotate=0] at (6.2,-2) {$\cal M$};
\node[rotate=0] at (11.2,5) {Time};

\end{tikzpicture}
    \caption{\small{Open path}} \label{fig:M1} 
  \end{subfigure}
  \qquad \qquad 
\begin{subfigure}[t]{0.4\linewidth}
    \begin{tikzpicture}[scale = 0.52]
    \useasboundingbox (-1,-3) rectangle (11,4.5);
\path[name path=border1] (0,0) to[out=-10,in=150] (6,-2);
\path[name path=border2] (12,1) to[out=150,in=-10] (5.5,3.2);
\path[name path=redline] (0,-0.4) -- (12,1.5);
\path[name intersections={of=border1 and redline,by={a}}];
\path[name intersections={of=border2 and redline,by={b}}];

\shade[left color=gray!10,right color=gray!80] 
  (-0.5,-0.5) to[out=-10,in=150] (6,-3) -- (12,1) to[out=150,in=-10] (5,3.7) -- cycle;
\draw (-0.3,3.5) to[out=-10,in=170] 
  coordinate[pos=0.27] (aux1) 
  coordinate[pos=0.72] (aux2) 
  coordinate[pos=0.75] (aux3) (10,5);
\coordinate (lambda_i) at (2,0.2);
\coordinate (lambda_f) at (9,1.2);
\coordinate (midpoint) at (6,0.1);

\node[below] at (lambda_i) {$\lambda_{\mathrm{i}} = \lambda_{\mathrm{f}} 
$};
\node[below right] at (lambda_f) {$\lambda(\ve t)$};

\tikzset{
    multiarrow/.style={
        decoration={markings,
            mark=at position 0.25 with {\arrow{>}},
            mark=at position 0.75 with {\arrow{>}}
        },
        postaction={decorate}
    }
}

\draw[line width=1.5pt, smooth, multiarrow, very thick, purple!70!black,fill=purple!50!white, fill opacity=0.3]
   (lambda_i) .. controls +(3,0.5) and +(0,1.0) .. (midpoint)
             .. controls +(-1,-3.8) and +(-0.5,0.8) .. (lambda_f) .. controls +(-6,3.8) and +(3.5,1.1) .. (lambda_i);

\draw[dashed] (aux1)to[bend left]  (lambda_i);
\draw[dashed] (aux2)to[bend left]  (lambda_f);

 \draw[fill=black] (lambda_i) circle (3pt);
 \draw[fill=black] (lambda_f) circle (3pt);


\foreach \coor in {1,2}
{
  \draw[fill=white] (aux\coor) circle (3pt);
}
\node[label=above:$0$] at (aux1) {};
\node[label=above:$t$] at (aux2) {};


\node[rotate=0] at (6.2,-2) {$\cal M$};
\node[purple!70!black] at (6,1.5) {$\mathscr{S}$};
\node[rotate=0] at (9.2,4.5) {Time};

\end{tikzpicture}
\caption{\small{Closed loop}} \label{fig:M2}  
\end{subfigure}
\vspace{-0.5 cm}
\caption{\small{Protocol $\Gamma = (\lambda(\ve t), 0\leq t\leq \tau/\ve)$ in parameter space $\cal M$. The closed loop $\Gamma = \partial \mathscr{S}$ traces the boundary of a surface $\mathscr{S}$.}}
\end{figure}

Suppose a physical system is exchanging energy with an equilibrium heat bath at temperature $T$. For its theoretical description, we write $\mathcal{M}$ for the parameter space, which can be taken as an $n-$dimensional differentiable manifold, but all examples in this manuscript take $\cal M \subset  \mathbb{R}^n$. For a vector $\lambda = (\lambda^1,..., \lambda^n) \in \mathcal{M}$, the entries $\lambda^{\mu}$ represent physical quantities such as environment temperature and/or pressure, internal agitation or flipping rates, {\it etc}. We take the  convention to write $\lambda = (\beta, \zeta)$ with $\zeta = (\zeta^1,..., \zeta^{n-1})$ 
to distinguish the inverse temperature of the bath $\beta = (k_B T)^{-1}$ from other parameters $\zeta$ appearing for instance in an energy function or as driving amplitudes. \\
It is of great importance to understand the response of the system when the parameters $\lambda $ are varied smoothly over a path $\Gamma$ in parameter space, called the protocol, say from some initial to a final value, $\lambda_{\text{i}} \rightarrow \lambda_{\text{f}} $. We assume that the variation is slow compared to the relevant relaxation times of the system. Mathematically, we ensure this by varying the parameters as $\lambda^{(\ve)}( t) = \lambda(\ve t)$ for $t \in [0, \ve^{-1} \tau]$ and macroscopic time $\tau$, as depicted in Fig.~\ref{fig:M1}. The parameter $\varepsilon$ is the rate of change along the protocol, and taking the limit $\varepsilon \downarrow 0$ makes the process quasistatic. The system dynamics itself, for all fixed parameters along the considered protocols, relaxes exponentially fast, uniformly in the initial condition.

\subsection{Quasistatic excess}\label{excess quantities intro}
The system observables of interest can be mathematically represented as (parameter-dependent) fluxes $f_t(\omega;\Gamma) = f_{\lambda(\ve t)}(\omega)$ on the system trajectories  $\omega = (X_t, 0\leq t\leq \tau/\ve)$ in a time window of size proportional to $\ve^{-1}$. The state $X_t$ denotes the mesoscopic condition of the nonequilibrium system at time $t$, for example, colloidal particle positions $x_t$ and velocities $v_t$, occupation of energy levels $E_t$, and spin configurations $\sigma_t$. At time zero, the system starts in a stationary condition where the state $X_0$ is sampled with probability density $\rho^{\mathrm{s}}_{\lambda_{\text{i}}}$, stationary for the dynamics running with parameters $\lambda_{\text{i}}$.\\
During the transformation, quantities change value, or extra currents may emerge.  Postponing further specification to Section \ref{marqu}, for now, we consider quite abstractly the total expected flux or current $J_{\lambda(\ve t)}(t) =  \int f_{\lambda(\ve t)}(\omega) \ \id \mathbb{P}(\omega)$, averaged over possible system trajectories $\omega$, and look at their time-integrated value over the protocol
\begin{equation}\label{Ha}
 H_{\ve}[f] = \int_0^{\tau/\varepsilon}\id t\, J_{\lambda(\ve t)}(t)
\end{equation}
For nonequilibrium purposes, the limit $\ve\downarrow 0$ of \eqref{Ha} often diverges, when at late times $t \simeq \tau/\ve \uparrow \infty $ the system reaches a new stationary condition with nonvanishing current. However, since we are primarily interested in the (integrated) flux {\it entirely} due to the change in parameter values $\lambda$, and not in the asympototically reached stationary flux, at every moment we want to subtract the instantaneous stationary current $J^{\mathrm{s}}_{\lambda(\ve t)}$, also known as the house-keeping part \cite{oono,jchemphys,Hatano_2001,MaesNetocny2015,Komatsu_2009}. That leads to the notion of a `renormalized' or {\it excess} current $H^{\text{exc}}$. Under quasistatic changes in parameters $\lambda$ along a path $\Gamma$ in parameter space, as depicted in Fig.~\ref{fig:M1}, we modify \eqref{Ha} into
\begin{align} 
    H_\ve^{\text{exc}}[f] & = \int_0^{\tau/\varepsilon}\id t \left(J_{\lambda(\ve t)}(t)-J^{\mathrm{s}}_{\lambda(\ve t)}\right) 
\end{align} 
leading to 
\begin{align} \label{definition excess H}
    H^{\text{exc}}(\Gamma) & = \lim_{\ve \downarrow 0} H_\ve^{\text{exc}}[f]
    = \lim_{\varepsilon \downarrow 0} \int_0^{\tau/\varepsilon} \id t \left(J_{\lambda(\ve t)}(t)-J^{\mathrm{s}}_{\lambda(\ve t)}\right) = \int_{\Gamma} \id \lambda \cdot R(\lambda).
\end{align}
Here $R(\lambda)$ is the quasistatic response function, which is of central interest. We refer to \cite{mathnernst,jchemphys} for mathematical details and for understanding $R(\lambda)$ as (nonequilibrium) heat capacity when $f_\lambda$ is the heat flux and the slow parameter is the temperature of a heat bath.\\

\noindent{\bf Remarks:}\\
Note that the limit $H^{\text{exc}}(\Gamma)$ is ``geometric,'' only depending on the curve $\Gamma$ in parameter space and not on how fast the parameters change in time, \textit{i.e.} invariant under a reparametrization of time $t\mapsto \gamma(t)$ for a smooth function $\gamma$ with $\gamma'(t)>0$. When we close the protocol
having a loop $\Gamma$ in parameter space, then $R(\lambda)$ becomes the equivalent of the Berry potential from which the Berry curvature $\Omega$ is derived; see Section \ref{max}.  We remind the reader that splitting of the total integrated current into a housekeeping and excess part is analogous to how in quantum mechanics with a time-dependent, cyclic Hamiltonian $H(t) = H_{\lambda(\ve t)}$ whose parameters vary slowly, the wave function picks up a phase $\phi_{\text{tot}}$ after one period $| \Psi(\cal T) \rangle = e^{i \phi_{\text{tot}}} |\Psi(0) \rangle = e^{i (\phi_\text{dyn} + \phi_B )} |\Psi(0) \rangle$ which splits into a dynamical part $\phi_{\text{dyn}} = \int_0^{\cal T}  \id t' \ \langle \Psi(t') | H_{\lambda(\ve t)} | \Psi(t') \rangle$ (housekeeping contribution) and the Berry phase $\phi_B  =  \oint   \id \lambda  \ \langle \Psi(\lambda) | \nabla_\lambda  | \Psi(\lambda) \rangle$ (excess part). \\
Secondly, we note that the response functions $R(\lambda)$ and Berry phase $H^{\text{exc}}$ typically also depend on other parameters that remain constant along the path $\Gamma$ (\textit{e.g.} in an isothermal process). \\  
Finally, under (global) detailed balance, the stationary currents $J_\lambda^{\mathrm{s}}$ in \eqref{definition excess H} typically vanish, and the excess current $H^{\text{exc}}$ reduces to the ordinary time-integrated flux \eqref{Ha}, and we recover the standard thermodynamic formalism. However, some excesses are real excesses also in equilibrium, so not for all choices of $f_\lambda$ do we have $J_\lambda^{\mathrm{s}} = 0$ automatically in equilibrium.  A natural observable with that property is the excess reactivity discussed in Example \ref{exre}. 

\vspace{-0.3 cm}

\subsection{Berry-ology for Markov jump processes}\label{marqu}
We move to the case where the description of the system coupled to a heat bath may proceed in terms of a Markov jump process $X_t$ for the system state at time $t$. We denote the states by $x, y, z, \dots \in K$ for the state space $K$.
The transition rate for the jump $x \rightarrow y$ is denoted by $k_\lambda(x, y)$ and it is zero iff no such jump is possible. The subscript $\lambda$ indicates the dependence of the transition rates on the parameters $\lambda$. 
The nonzero rates $k_\lambda(x,y)$ can always be decomposed as 
\begin{equation}\label{dec}
k_\lambda(x,y) = a_\lambda(x,y) \ e^{s_\lambda(x,y)/2}, \qquad a_\lambda(x,y)=a_\lambda(y,x),\qquad s_\lambda(x,y)=-s_\lambda(y,x),
\end{equation}
where, under the exchange of (edge direction) $x\leftrightarrow y$,  the edge reactivity $a_\lambda$ is symmetric and  $s_\lambda$ is antisymmetric, 
\begin{align}
    a_\lambda(x,y) = \sqrt{k_\lambda(x,y) \ k_\lambda(y,x)}, \qquad     s_\lambda(x,y) = \log \left( \frac{k_\lambda(x,y)}{k_\lambda(y,x)} \right) \label{a = sqrt k}
\end{align}

Under local detailed balance \cite{ldb,frenesy}, it is possible to interpret
\begin{align}\label{ldb}
    q_\lambda(x,y) = \frac{1}{\beta} \log \left( \frac{k_\lambda(x,y)}{k_\lambda(y,x)} \right) = k_B T\,s_\lambda(x,y)
\end{align}
as the heat $q_\lambda(x, y) = -q_\lambda(y, x)$ released to the bath during $x \to  y$.  Then, $s_\lambda(x,y)$ is the entropy change (per $k_B$) in the thermal bath from the transition $x \to y$. \\

It is convenient to work with the backward generator  $L_\lambda$ of  the process $X_t$, defined on real-valued functions $g$ on $K$ by
\begin{equation}\label{lap}
L_\lambda g(x) = \sum_{y \in K}k_\lambda(x, y)[g(y) - g(x)]    
\end{equation}
and 
$e^{t L_\lambda} g(x) = \langle g(X_t) | X_0 = x\rangle_\lambda$.
The forward generator $L^\dagger_\lambda$ is  the transpose of $L_\lambda$, meaning that
\begin{align}\label{defadjoint}
\sum_{x \in K} L_\lambda b(x)\, c(x) = \sum_{x \in K} b(x)\, L^\dagger_\lambda c(x) 
\end{align}
for all functions $b,c$. Observe that always $\sum_{x \in K} L^\dagger_\lambda c(x) =0$ for every function $c$.  The stationary probability distribution is denoted by $\rho_\lambda^{\mathrm{s}}$ and satisfies $L^\dagger_\lambda \rho_\lambda^{\mathrm{s}}(x) = 0$ for each $x$. \\

Imagine next that the transition rates are time-dependent via a quasistatic protocol $\lambda^{(\ve)}( t) = \lambda(\ve t)$. We denote by $\rho_t^\ve$ the solution to the time-dependent Master equation 
\begin{align}\label{time dependent master eq}
\frac{\partial \rho^\ve_{t}}{\partial t}(x) &= L^\dagger_{\lambda(\ve t)} \rho^\ve_{t}(x) = \sum_{y \in K}[\rho^\ve_{t}(y) \ k_{\lambda(\ve t)}(y,x) - \rho^\ve_{t}(x) \ k_{\lambda(\ve t)}(x,y)]
\end{align}
with normalization $\sum_{x \in K} \rho_t^\ve(x) = 1.$ We also define the averages $\langle \cdot \rangle_t^\ve$ and $\langle \cdot \rangle_\lambda^{\mathrm{s}}$,
\begin{align}
 \qquad  \langle g \rangle_t^\ve = \sum_{x \in K} g(x) \ \rho_t^\ve(x),\qquad \langle g \rangle_\lambda^{\mathrm{s}} = \sum_{x \in K} g(x) \ \rho_\lambda^{\mathrm{s}}(x).
\end{align}

Thinking now about the temporal excesses as in \eqref{Ha}--\eqref{definition excess H}, we are interested in single-time observables $f_\lambda$ expressing an expected rate of change or flux, of the form 
\begin{equation}\label{re}
f_\lambda(x) = \sum_{y \in K} k_\lambda(x,y) \ h_\lambda(x,y),
\end{equation}

where we sum over all possible $y \in K$ to account for all possible transitions from $x$, assuming only 1 jump can occur in a small enough $\id t$. 
Note that $f_\lambda$ indeed takes the interpretation of a flux, since the transition rate $k_\lambda(x,y)$ has the unit of frequency. What the flux physically represents depends on the function $h_\lambda(x,y)$, and some  cases of physical interest correspond to
\begin{itemize}
\item Heat flux $\cal P_\lambda$:
\begin{align}
    h_\lambda(x,y) = q_\lambda(x,y) \Longrightarrow f_\lambda(x) = \cal P_\lambda(x) := \sum_{y \in K} k_\lambda(x,y) \ q_\lambda(x,y) \label{heat flux}
\end{align}
    \item Entropy flux $\Sigma_\lambda$:
    \begin{align}
        h_\lambda(x,y) = s_\lambda(x,y) \Longrightarrow f_\lambda(x) = \Sigma_\lambda(x) : =  \sum_{y \in K} k_\lambda(x,y) \ s_\lambda(x,y) \label{rent} 
    \end{align}
    \item (State) reactivity $A_\lambda$:
    \begin{align}
        h_\lambda(x,y) = e^{-s_\lambda(x,y)/2} \Longrightarrow f_\lambda(x) = A_\lambda(x) := \sum_{y \in K} a_\lambda(x,y)\label{activity}
    \end{align}
    \item Escape rate $\xi_\lambda$:
    \begin{align}
        h_\lambda(x,y) = 1 \Longrightarrow f_\lambda(x) = \xi_\lambda(x) := \sum_{y \in K} k_\lambda(x,y) \label{escape rate}
    \end{align}
    \item Transition rate $k_\lambda$:
    \begin{align}
        h_\lambda(x,y) = \delta_{y, y'} \Longrightarrow f_\lambda(x) = k_\lambda(x,y') \label{f = k}
    \end{align}
     for some fixed $y' \in K$, relevant for excess currents.
\item Rate of change of state function:
    \begin{align}\label{h state funtion}
        h_\lambda(x,y) = \tilde{h}_\lambda(x)-\tilde{h}_\lambda(y) \Longrightarrow f_\lambda(x) = \sum_{y \in K} k_\lambda(x,y) \left[\tilde{h}_\lambda(x)-\tilde{h}_\lambda(y) \right] = -L_\lambda \tilde{h}_\lambda(x),
    \end{align}
where we used \eqref{lap}. A particular case of interest is the energy function $\tilde{h}_\lambda(x) = E_\zeta(x)$.
\end{itemize}

 For the physical meaning of the state reactivity $A_\lambda(x) = \sum_{y \in K} a_\lambda(x,y)$ in \eqref{activity}, one remembers from \eqref{a = sqrt k} that the edge reactivity $a_\lambda(x,y)$ is a transition frequency, independent of the direction of the flow. This observable represents a real excess in equilibrium as well, since, under the Boltzmann-Gibbs equilibrium distribution $\rho_\lambda^{\text{eq}}$, there is no need for $\left \langle A_\lambda \right \rangle_\lambda^{\text{eq}} $ to vanish. \\ 

For no matter what $h_\lambda(x,y)$ in \eqref{re}, the time-integrated current in \eqref{Ha} becomes
\begin{align}\label{current in rho_t^eps}
    J_{\lambda(\ve t)}(t) & = \langle f_{\lambda(\ve t)} \rangle_t^\ve =  \sum_{y \in K} f_{\lambda(\ve t)}(x) \  \rho_t^\ve(x), \qquad J^{\mathrm{s}}_{\lambda}  = \langle f_{\lambda} \rangle^{\mathrm{s}}_\lambda = \sum_{y \in K} f_\lambda(x) \  \rho_\lambda^{\mathrm{s}}(x),
\end{align}
with time-accumulated excess \eqref{definition excess H},
\begin{align}\label{excess with f}
      H_\ve^{\text{exc}}[f]
      =  \int_0^{\tau/\ve} \id t \, \sum_{x \in K} f_{\lambda(\ve t)}(x)\,\left(\rho_t^\ve(x) - \rho^{\mathrm{s}}_{\lambda(\ve t)}(x) \right),
\end{align}
where $\lambda(\ve t)$ varies over the curve $\Gamma$, taking a macroscopic time $\tau$. To compute $H_\ve^{\text{exc}}[f]$ in the limit $\ve \downarrow 0$, assuming everything is smooth, we use\footnote{The inverse $1/L_\lambda^\dagger $ is defined on vectors $g$ having zero sum, $\sum_{x \in K} g(x)=0$ (like $g=\nabla_\lambda\rho_\lambda^{\mathrm{s}}$), and $\frac1{L_\lambda^\dagger} g = \cal V_\lambda$  is the unique solution of the (forward) Poisson equation,
\begin{align*}
L^\dagger_\la \cal V\, (x) = g(x),\qquad \sum_{x \in K} \cal V(x)=0.
\end{align*}} the result, \cite{mathnernst,jchemphys},
\begin{align}
\lim_{\ve \downarrow 0}\int_0^{ \tau/\ve}(\rho^\ve_t - \rho^{\mathrm{s}}_{\lambda(\ve t)}) \ \id t &= \int_\Gamma\id \lambda\cdot \frac 1{L_\lambda^\dagger}\nabla_\lambda\rho_\lambda^{\mathrm{s}} \label{mm} 
\end{align}
Therefore,
\begin{align}
\lim_{\ve \downarrow 0}\int_0^{ \tau/\ve} \sum_{x \in K} f_{\lambda(\ve t)}(x) \ (\rho^\ve_t(x) - \rho^{\mathrm{s}}_{\lambda(\ve t)}(x)) \ \id t &= \int_\Gamma\id \lambda\cdot \sum_{x \in K} (f_{\lambda}(x) - \langle f_\lambda\rangle^{\mathrm{s}}_\lambda) \frac 1{L_\lambda^\dagger}\nabla_\lambda\rho_\lambda^{\mathrm{s}}(x), \label{mm2}
\end{align}
where we can insert the stationary average $\langle f_\lambda \rangle_\lambda^{\mathrm{s}}$ since $\sum_{x \in K} \frac{1}{L_\lambda^\dagger} \nabla_\lambda \rho_\lambda^{\mathrm{s}}(x) = 0$. 
The last integral is over the components $\lambda^\mu$ of the vector $\lambda$, which represent the moving parameters of our choice, possibly including the temperature. 
Using \eqref{defadjoint}, the limit \eqref{mm2} can be written as, \cite{mathnernst,jchemphys},
\begin{align}
    \sum_{x \in K} \left(f_{\lambda}(x) -  \langle f_\lambda\rangle_\lambda^{\mathrm{s}}\right) \big(\frac 1{L_\lambda^\dagger} \nabla_\lambda\rho_\lambda^{\mathrm{s}} \big)(x) &=  \sum_{x \in K} \frac 1{L_\lambda} (f_\lambda  - \langle f_\lambda\rangle_\lambda^{\mathrm{s}})(x) \ \nabla_\lambda\rho_\lambda^{\mathrm{s}}(x), \\
-\frac 1{L_\lambda} (f_\lambda - \langle f_\lambda\rangle_\lambda^{\mathrm{s}})(x)=    V_\lambda(x) & = \int_0^\infty \id t \ e^{t L_\lambda} \left[f_\lambda(x) - \left \langle f_\lambda\right \rangle_\lambda^{\mathrm{s}} \right], \label{def Vlam}
\end{align}
which also defines $1/L_\la$.
Here, $V_\lambda$ is called the quasipotential representing the unique solution to 
the Poisson equation\footnote{The name ``Poisson'' of the equation comes from viewing $L_\lambda$ as the (graph) Laplacian; see also \cite{drazin,pois}.  The terminology ``quasipotential'' is less standard, and refers here to the solution of such a Poisson equation, in perfect harmony with its origin from quasistatic transformations.} on $K$ sourced by $f_\lambda$:
\begin{align}\label{Pois}
    L_\lambda V_\lambda (x) + f_\lambda(x) - \langle  f_\lambda\rangle_\lambda^{\mathrm{s}} =0, \qquad \langle V_\lambda \rangle_\lambda^{\mathrm{s}} = 0.
\end{align}
Via \eqref{mm2}--\eqref{Pois}, the excess \eqref{definition excess H} then becomes
\begin{equation}
    H^{\text{exc}}  = \lim_{\ve \downarrow 0}  H_\ve^{\text{exc}}
    = \int_{\Gamma} \id \lambda \cdot  \left \langle \nabla_\lambda V_\lambda \right \rangle_\lambda^{\mathrm{s}}  =  - \int_\Gamma\id \lambda\cdot \left \langle V_{\lambda}
    \, \nabla_\lambda \log \rho^{\mathrm{s}}_\lambda \right \rangle_\lambda^{\mathrm{s}}\label{limit excess H}
\end{equation}
Comparing \eqref{limit excess H} to \eqref{definition excess H}, we recognize 
the response coefficient
    \begin{equation}\label{ala}
       R(\lambda) =  \left \langle \nabla_\lambda V_\lambda \right \rangle_\lambda^{\mathrm{s}} = - \left \langle V_{\lambda}
    \, \nabla_\lambda \log \rho^{\mathrm{s}}_\lambda \right \rangle_\lambda^{\mathrm{s}},
    \end{equation}
    where the equality holds since $\langle V_\lambda\rangle_\lambda^{\mathrm{s}}=0  $. We emphasize that $V_\lambda$ obviously depends on the choice of $h_\lambda$ in \eqref{re}, but that dependence on the physical flux under consideration is ignored in the notation.\\
Remark that the presence of $\nabla_\la \rho_\lambda^{\mathrm{s}}$ in the above equations, starting with \eqref{mm}, reminds us again of the expression of the Berry phase in quantum mechanics  $\phi_B  =  \oint   \id \lambda  \ \langle \Psi(\lambda) | \nabla_\lambda  | \Psi(\lambda) \rangle$ involving the
gradient of a wave function. That connection is not very surprising as geometric aspects on Floquet theory are basically unchanged when moving to jump processes \cite{Schindler_2025}. Moreover, a formal connection may appear from the Doi–Peliti formalism \cite{peliti1985path, AltlandSimons2023}, where jump processes are mapped to a Fock-space representation using creation and annihilation operators, resembling the quantum formalism.\\
A second remark concerns possible gauge dependence. In the quantum context, the wave function has an arbitrary phase degree of freedom $|\Psi(\lambda) \rangle \to |\Psi'(\lambda) \rangle = e^{i \chi(\lambda)}| \Psi(\lambda) \rangle$ which does not change the physical properties of the state (gauge transformation), resulting in a change in the gauge-dependent quantum Berry potential $\mathbf{A}(\lambda) \to \mathbf{A}'(\lambda) = \mathbf{A}(\lambda) + i\nabla_\lambda \chi(\lambda)$. Similarly, without centering the quasipotential,  $V_\lambda$ as the solution to \eqref{Pois} is not unique since one can always add to $V_\lambda$ a $\lambda$-dependent constant $C_\lambda$ (not depending on the state $x$) for which $L_\lambda C_\lambda = 0$ (gauge freedom). That results in a change of the Berry potential $R(\lambda)$
    \begin{align*}
        R(\lambda) \to R'(\lambda) = R(\lambda) + \nabla_\lambda C_\lambda
    \end{align*}
    similar to the quantum mechanical case. However, as was done \eqref{Pois}, one typically chooses, with a clear thermodynamic interpretation, to center the quasipotential $\langle V_\lambda \rangle_\lambda^{\mathrm{s}} = 0$, which can be seen as a choice of gauge from this perspective. The reason for doing so is that our description is not microscopic, but mixes mesoscopic dynamics with thermodynamic concepts, and we believe to {\it know} exactly what the physical {\it real} energies, currents, and excesses are, which leaves no room for a gauge freedom in $V_\lambda$. In that context, we also refer to \cite{Pe_2012}, where, for Markov jump processes, a gauge-dependence in the energy and work function is discussed, keeping the heat fixed.\\

Alternatively, for a given edge $(x,y)$ and transition rate $k_\lambda(x,y)$ as in \eqref{f = k}, we get the excess (time-integrated) current by taking the difference
\begin{equation}\label{exccu}
     \cal  J_\ve^{\text{exc}}(x,y)=\int_{0}^{\tau/ \ve} \id t \left[ k_{\lambda (\ve t)}(x,y) \big(\rho^\ve _t (x) - \rho^{\mathrm{s}}_{\lambda(\ve t)}(x)\big)- k_{\lambda (\ve t)}(y,x)\big(\rho^\ve _t (y) - \rho^{\mathrm{s}}_{\lambda(\ve t)}(y)\big)\right].
\end{equation}
As we integrate over time, what we call ``excess current'' is really the excess in the net number of jumps across the edge $x \to y$.  In the quasistatic limit where $\ve \downarrow 0$ over a path $\Gamma$,  we obtain using \eqref{mm}
\begin{align}
    \mathcal{J}^{\text{exc}}(x,y) = \lim_{\ve \downarrow 0} \cal J_\ve^{\text{exc}}(x,y)
    &= \int_{\Gamma} \id \lambda \cdot   k_{\lambda}(x,y)\frac{1}{L^{\dagger}_{\lambda}} \nabla_{\lambda} \rho_{\lambda}^{\mathrm{s}}(x)\, 
    - \int_{\Gamma} \id \lambda \cdot \, k_{\lambda}(y,x) \frac{1}{L^{\dagger}_{\lambda}} \nabla_{\lambda} \rho_{\lambda}^{\mathrm{s}}(y)\notag\\
    &=\int_{\Gamma} \id \lambda \cdot  [ k_{\lambda}(x,y) \ \cal V_\lambda(x) - \, k_{\lambda}(y,x) \ \cal V_\lambda(y)],
    \label{jint}
\end{align}
where $\cal V_\la$ is the (vector) solution
of the forward Poisson equation, 
\begin{equation}
 L^{\dagger}_\lambda \cal V_{\lambda} (x) = \sum_{y \in K} k(y,x) \mathcal{V}_{\lambda}(y) - k(x,y) \mathcal{V}_{\lambda}(x) = \nabla_{\lambda} \rho_{\lambda}^{\mathrm{s}}(x)
 \label{poiscu1}
\end{equation}
subject to the condition $\sum_{x \in K} \mathcal{V}_{\lambda}(x) = 0$.

Such an excess (geometric) current \eqref{jint} has been calculated in \cite{Sinitsyn_2007} for a double--channel two-state system; we give another illustration in Example \ref{excessCurrent}. \\

Formula \eqref{limit excess H} is our general point of departure. To connect this excess with the Berry phase, \cite{Berry1984, Berry_1985, wilczek1989geometric, hannay, Kariyado_2016, harrythirty, Sinitsyn_2007}, we consider for $\Gamma$ a closed path in parameter space that traces out the boundary $\partial \mathscr{S}$ of a surface $\mathscr{S}$ with a positive orientation (counterclockwise) as depicted in Fig. \ref{fig:M2}.
The expression $H^\text{exc}$ in \eqref{limit excess H} then becomes the ``Berry phase''  and the response function gets interpreted as the ``Berry potential'' $R(\lambda)$,
\begin{align}
   H^{\text{exc}} =  \oint_\Gamma   \id \lambda  \cdot R(\lambda) = \oint_\Gamma \omega, \qquad  R(\lambda) = \frac{\delta H^{\text{exc}}}{\id \lambda } 
   =  -\sum_{x \in K}  \ V_\lambda(x) \  \nabla_\lambda \rho_{\lambda}^{\mathrm{s}}(x) \label{bpha}
\end{align}
with one form $\omega = R_\mu \ \id \lambda^\mu$ where the sum over $\mu$ is implied. 
Here $R_\mu$ represents the $\mu$th component of the vector $R(\lambda)$ and is given by
\begin{align}\label{A components}
R_\mu = - \sum_{x \in K}  \ V_\lambda(x) \ \partial_\mu  \rho_{\lambda}^{\mathrm{s}}(x) = \sum_{x \in K} \partial_\mu V_\lambda(x) \ \rho_\lambda^{\mathrm{s}}(x) = \left \langle  \partial_\mu V_\lambda \right \rangle_\lambda^{\mathrm{s}}, \qquad  \partial_\mu = \frac{\partial}{\partial \lambda^\mu} 
\end{align}
Note that $V_\la$ solves the Poisson equation \eqref{Pois} sourced by $f_\la$.
Therefore, we can also write the response vector (or Berry potential) as
\begin{align}
R(\la) = - \sum_{x \in K}  \ V_\lambda(x) \ L^\dagger_\la \cal V_\la(x) = - \sum_{x\in K} f_\la(x), \,\cal V_\la(x)
  \end{align}
where $\cal V_\la$ with $\sum_x \cal V_\la(x)=0$ solves \eqref{poiscu1}.\\

We end this section with physically interesting examples of excesses and corresponding responses (Berry potentials).
The corresponding Berry curvature is discussed in the following sections.

\begin{example}[Thermal response]\label{thre}
    Taking the heat flux \eqref{heat flux}, we obtain the excess heat $ H_\ve^{\text{exc}} =  Q_\ve^{\text{exc}} $ and quasipotential $V_\lambda^\cal P$,
\begin{align}
    Q_\ve^{\text{exc}}& = \int_0^{\tau/\ve} \id t \sum_{x \in K}\cal P_{\lambda(\ve t)}(x) \left(\rho_t^\ve(x) - \rho^{\mathrm{s}}_{\lambda(\ve t)} \right), \nonumber \\
    Q^{\text{exc}} &= \lim_{\ve \downarrow 0}  Q_\ve^{\text{exc}} = \int_{\Gamma} \id \lambda \cdot  \left \langle \nabla_\lambda V_\lambda^{\cal P} \right \rangle_\lambda^{\mathrm{s}} =- \int_\Gamma \id  \lambda\cdot \left \langle V_{\lambda}^{\cal P}
    \, \nabla_\lambda \log \rho^{\mathrm{s}}_\lambda \right \rangle_\lambda^{\mathrm{s}}\label{limit excess heat}, \\
    V^\cal P_\lambda(x)& = \int_0^\infty \id t \ e^{t L_\lambda} \left[\mathcal P_\lambda(x) - \left \langle \mathcal P_\lambda\right \rangle_\lambda^{\mathrm{s}} \right], \label{VP} 
\end{align}
where we add a superscript $\cal P$ on $V_\lambda$ to distinguish the different quasipotentials in the examples. When the only parameter that is varied is the temperature, $T \to T + \id T$, we get that the thermal response function $R_\beta^{\cal P}(\lambda)$ is directly related to the (nonequilibrium) heat capacity $C^{\text{neq}}_\lambda$ which is the excess heat produced per $\id T$, \cite{jchemphys,epl,calo}:
\begin{align}\label{HCReponse}
 R_\beta^{\cal P}(\lambda) = \frac{\delta  Q^{\text{exc}}}{\id \beta}, \qquad  \; C^{\text{neq}}_\lambda =  -\frac{\delta  Q^{\text{exc}}}{\id T} = \beta^2\frac{\delta  Q^{\text{exc}}}{\id \beta} = \beta^2 \left \langle \frac{\partial V_\lambda^{\cal P}}{\partial \beta} \right \rangle_\lambda^{\mathrm{s}}.
\end{align}
We extend this example to the case of two thermal baths by taking parameters $\lambda = (\beta_1,\beta_2,\zeta)$, specified as follows. We consider the simple setup of a three-level system with states $x,y,z$ and energy gaps $\delta_1, \delta_2$
in \fig\ref{ziaex}. The ``molecule'' is in contact with two thermal baths at inverse temperatures $\beta_1 $ and $\beta_2$, representing a toy model of a  (molecular) conductor.
 Finally, there are transitions between $x$ and $y$ with constant switching rate $\alpha$.
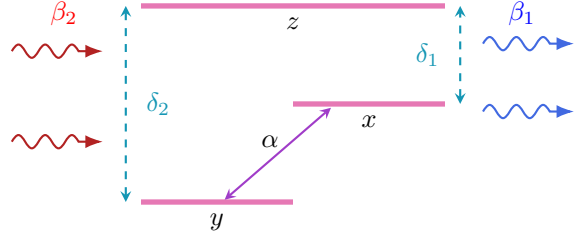
\begin{figure}[H]
\begin{tikzpicture}[scale=1,>=stealth]

  \tikzset{
     heatL/.style={FireBrick,thick,-Latex, decoration={coil,aspect=0},decorate},
     heatR/.style={RoyalBlue,thick,-Latex, decoration={coil,aspect=0},decorate}
  }

  \foreach \x in {0.6,1.8} {
    \draw[heatL] (-1.7,\x+0.2) -- ++(1.2,0);
  }

  \foreach \x in {1.0,1.9} {
    \draw[heatR] (4.5,\x+0.2) -- ++(1.2,0);
  }

  \definecolor{levelcolor}{RGB}{230,120,180}
  \definecolor{arrowcolor}{RGB}{150,60,180}
  \definecolor{energycolor}{RGB}{20,150,180}
  \definecolor{photoncolor}{RGB}{180,40,40}
  \definecolor{purpleline}{RGB}{160,60,200}

  \draw[line width=2pt,levelcolor] (0,0) -- (2,0) node[midway, below,black] {$y$};
  \draw[line width=2pt,levelcolor] (2,1.3) -- (4,1.3) node[midway, below,black] {$x$};
  \draw[line width=2pt,levelcolor] (0,2.6) -- (4,2.6) node[midway, below,black] {$z$};

      \node at (1.7,0.8) {$\alpha$};


  \draw[energycolor,<->,thick,dashed] (-0.2,0) -- (-0.2,2.6)
    node[midway,right=4pt] {$\delta_2$};
  \draw[energycolor,<->,thick,dashed] (4.2,1.3) -- (4.2,2.6)
    node[midway,left=4pt] {$\delta_1$};

  \draw[purpleline, thick,<->] (1.1,0.03) -- (2.5,1.25)
    node[midway,sloped,above,black] {};

  \node[red,left=6pt] at (-0.5,2.5) {$\beta_2$};
  \node[blue,right=6pt] at (4.5,2.5) {$\beta_1$};

\end{tikzpicture}
\centering
\caption{\small{Heat conducting three-level system with energy gaps $\delta_1, \delta_2$.  The transitions between $(z,y)$ and between $(x,z)$ are thermal from left and right heat baths at inverse temperatures $\beta_1, \beta_2 $.
The externally applied switch $x\leftrightarrow y$ has constant rate $\alpha$. 
}}
 \label{ziaex}
 \end{figure}

 The transition rates in Fig.~\ref{ziaex} are
 \begin{align}\label{MCrate}
           k_\lambda(z,y)&=k_\lambda(z,x)=1, \qquad k_\lambda(x,z)=e^{-\beta_1 \delta_1}\notag \\
           k_\lambda(y,z)&=e^{-\beta_2 \delta_2}, \qquad k_\lambda(x,y)=k_\lambda(y,x)=\alpha\, 
\end{align}
Solving the stationary Master equation, the stationary distribution $\rho^{\mathrm{s}}_\lambda$ is found,
\begin{align*}
    \rho^{\mathrm{s}}_\lambda(x)&=\frac{e^{\beta _1 \delta _1}}{\cal Z} \left(2 \alpha  e^{\beta _2 \delta_2}+1\right), \qquad \rho^{\mathrm{s}}_\lambda(y)= \frac{e^{\beta _2 \delta _2}}{\cal Z} \left(2 \alpha  e^{\beta _1 \delta _1}+1\right) \\
    \rho^{\mathrm{s}}_\lambda(z)&=\frac{1}{\cal Z} \left(\alpha  (e^{\beta _1 \delta _1}+e^{\beta _2 \delta _2})+1\right),
\end{align*} 
where  $\cal Z=(\alpha +1) e^{\beta _1 \delta _1}+e^{\beta _2 \delta _2} \left(4 \alpha  e^{\beta _1 \delta _1}+\alpha +1\right)+1$. When $\beta_1 = \beta_2 =\beta$ and $\alpha = 0$, we get the usual equilibrium distribution with  $\cal Z=  e^{\beta \delta _1}+e^{\beta \delta _2} +1$. 
The expected heat fluxes for every state into the two baths are obtained by \eqref{heat flux},
\begin{align*}
    \cal P_1(x)&= -\delta _1 e^{- \beta _1 \delta _1}, \qquad \cal P_1(y)= 0, \qquad \cal P_1(z)=\delta_1  \\
   \cal P_2(x)&=0, \qquad  \cal P_2(y)= -\delta _2 e^{- \beta _2 \delta _2}, \qquad \cal P_2(z)=\delta_2.
\end{align*}
There is a matrix of heat capacities with elements
\begin{align}\label{mC}
C_{\beta_i\, \beta_j} = \beta_j^{\,2}\, \left\langle \frac{\partial V_\lambda^{\mathcal P_{i}}}{\partial \beta_j} \right\rangle_\lambda^{\mathrm{s}},
\end{align}
where $\cal P_i$ refers to the heat flux released to the $i$-th bath.  The diagonal elements of $C$ correspond to the heat released to a bath when its own temperature changes, while for $i\not =j$, the heat released to bath $i$ as a result of changing the temperature of bath $j$ is of interest.
\begin{figure}[H]
    \centering
    \begin{subfigure}{0.49\textwidth}
        \centering
        \includegraphics[scale=0.8]{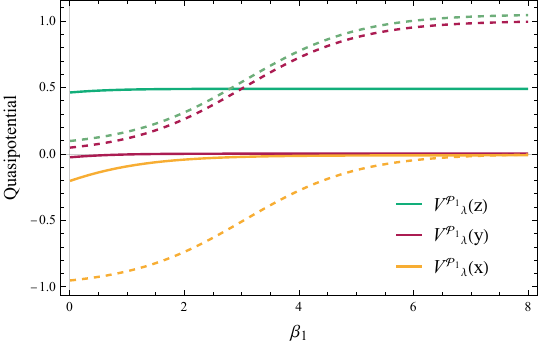}
        \caption{\small{$\beta_2=1.5$.}}
    \end{subfigure}
    \hfill
    \begin{subfigure}{0.49\textwidth}
        \centering
        \includegraphics[scale=0.8]{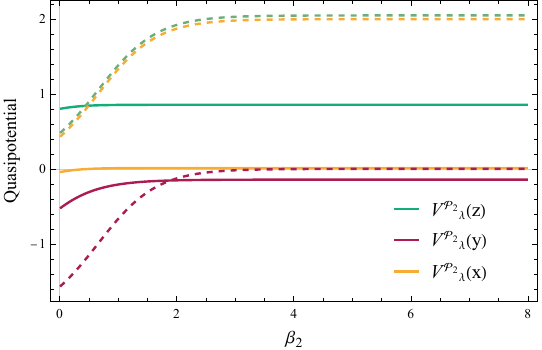}
        \caption{\small{$\beta_1=1$.}}
    \end{subfigure}
    \caption{\small{The quasipotentials $V^{\cal P_i}_\lambda$ (used in \eqref{mC}) of the heat flux as a function of the inverse temperatures with $\delta_1=1, \delta_2=2$. The solid lines have $\alpha=1$ and the dashed lines correspond to $\alpha=0$, keeping the same colors. We work here, and in the rest of the manuscript, with dimensionless units. }}\label{quasiheatMC}
\end{figure}

The elements of that heat capacity matrix are obtained as 
\begin{align*}
C_{\beta_1\, \beta_1} =&\frac{1}{\cal Z^3}\beta _1^2 \delta _1^2 e^{\beta _1 \delta _1} \left(2 \alpha  e^{\beta _2 \delta _2}+1\right) \\
&\times  \bigg[e^{2 \beta _2 \delta _2}\left(\alpha ^2 \left(4 e^{\beta _1 \delta _1}+2\right)+\alpha +1\right)+e^{\beta _2 \delta _2} \left((3 \alpha +1) e^{\beta _1 \delta _1}+2 (\alpha +1)\right)+e^{\beta _1 \delta _1}+1\bigg],\\
C_{\beta_1\, \beta_2} =&\frac{1}{\cal Z^3}e^{\beta _2 \delta _2} \beta _2^2 \delta _1 \delta _2 \left(2 \alpha  e^{\beta _1 \delta _1}+1\right) \\
&\times \bigg[(\alpha -1) e^{\beta _1 \delta _1} \left(e^{\beta _1 \delta _1}+1\right)+e^{\beta _2 \delta _2} \left(e^{\beta _1 \delta _1} \left[\alpha^2  \left(4 e^{\beta _1 \delta _1}+2\right)+3 \alpha -1\right]+\alpha \right)\bigg].
\end{align*}
By symmetry, exchanging $\beta_1 \leftrightarrow \beta_2$ and $\delta_1 \leftrightarrow \delta_2$, yields the other components $C_{\beta_2\, \beta_2}$ and $C_{\beta_2\, \beta_1}$.  From \fig \ref{quasiheatMC}, we observe that the quasipotentials from which these thermal responses are derived using \eqref{mC} remain bounded at low temperatures. We show in Section \ref{section extended third law} that this property is an essential condition for the vanishing of the Berry potential and curvature at zero temperature. And indeed, all heat capacities \eqref{mC} vanish (exponentially fast) for $\beta_1,\beta_2\uparrow \infty$, as a nonequilibrium example of the Nernst postulate, \cite{jchemphys}.\\

One can also look at the dependence of the excess heat absorbed from one specific heat bath on changing the switching rate $\alpha$, and compute $R^{\cal P_i}_\alpha$,
\begin{align*}
    R^{\cal P_1}_\alpha&= \left \langle \frac{\partial V_\lambda^{\cal P_1}}{\partial \alpha}\right \rangle_\lambda^{\mathrm{s}}=-\frac{1}{\cal Z^3}\delta _1 (e^{\beta _1 \delta _1}-e^{\beta _2 \delta _2})\bigg[-2 (\alpha -1) e^{\beta _1 \delta _1+2 \beta _2 \delta _2}-\alpha  e^{2 \beta _2 \delta _2}\\
    &\qquad \qquad \qquad \qquad \qquad \qquad +2 (\alpha +1) e^{2 \beta _1 \delta _1+\beta _2 \delta _2}+(\alpha +3) e^{\beta _1 \delta _1+\beta _2 \delta _2}+e^{\beta _1 \delta _1}+e^{2 \beta _1 \delta _1}\bigg],\\
    R^{\cal P_2}_\alpha&= \left \langle \frac{\partial V_\lambda^{\cal P_2}}{\partial \alpha}\right \rangle_\lambda^{\mathrm{s}}=- \frac{1}{\cal Z^3} \delta _2 (e^{\beta _2 \delta _2}-e^{\beta _1 \delta _1}) \bigg[-e^{2 \beta _1 \delta _1} (2 (\alpha -1) e^{\beta _2 \delta _2}+\alpha ) \\
    &\qquad \qquad \qquad \qquad \qquad \qquad +e^{\beta _1 \delta _1+\beta _2 \delta _2} (2 (\alpha +1) e^{\beta _2 \delta _2}+\alpha +3)+e^{\beta _2 \delta _2}+e^{2 \beta _2 \delta _2}\bigg].
\end{align*}
\fig~\ref{onlybetaMC} shows these responses as function of $\beta_2$ for $\alpha=1, \beta_1 = 0.5$ and $\beta_1 = 3$.  
\begin{figure}[H]
    \centering
    \includegraphics[width=0.5\linewidth]{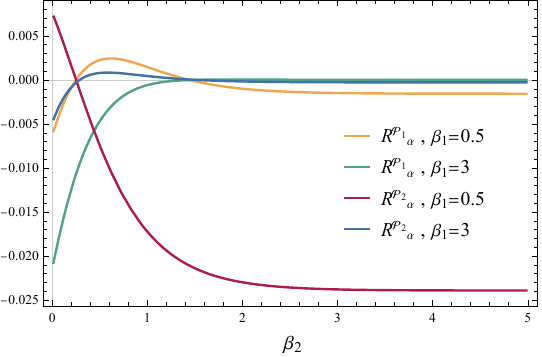}
    \caption{\small{The response $ R_\alpha^{\cal P_i}$ of the excess heat to each of the heat baths upon changing the switching rate $\alpha$ in Fig.~\ref{ziaex} around $\alpha=1$. We see the dependence on the bath temperatures for $\delta_1=1, \delta_2=2$.}}
    \label{onlybetaMC}
\end{figure}
A further analysis of this example is presented in Section~\ref{section extended third law}. 
\end{example}

\begin{example}[Excess reactivity]\label{exre}

For jump processes, dynamical activity refers to time-symmetric traffic over one or more edges.   Here, following the parameterization \eqref{dec}, we take the excess state reactivity \eqref{activity} with corresponding current
\begin{align*}
     J_{\lambda(\ve t)}(t) & =  \sum_{x \in K} A_{\lambda(\ve t)}(x) \  \rho_t^\ve(x) 
\end{align*}
 which measures a weighted expected rate of jumps per unit time under the distribution $\rho_t^\ve$.  Its excess 
 \[
 H_\ve^{\text{exc}} =  \mathcal{A}_\ve^{\text{exc}} =  \sum_{x\in K}\int_0^{\tau/\ve} \id t \ A_{\lambda(\ve t)}(x) \left(\rho_t^\ve(x) - \rho^{\mathrm{s}}_{\lambda(\ve t)} \right)  \nonumber \\
 \]
 has the limit
\begin{align}
    \mathcal{A}^{\text{exc}} &= \lim_{\ve \downarrow 0}  \mathcal{A}_\ve^{\text{exc}} = \int_{\Gamma} \id \lambda \cdot  \left \langle \nabla_\lambda V_\lambda^{\cal{A}} \right \rangle_\lambda^{\mathrm{s}}  =- \int_\Gamma \id  \lambda\cdot \left \langle V_{\lambda}^{\cal{A}}
    \, \nabla_\lambda \log \rho^{\mathrm{s}}_\lambda \right \rangle_\lambda^{\mathrm{s}}\label{limit excess dynamical activity} \\
    V^\cal{A}_\lambda(x) &=  \int_0^\infty \id t \ e^{t L_\lambda} \left[A_\lambda(x)- \left \langle A_\lambda\right \rangle_\lambda^{\mathrm{s}} \right]. \label{VA} 
\end{align}
To give a specific example, we calculate the quasipotential $V_\lambda^A$ of the excess reactivity for a  Markov jumper on the graph in  \fig\ref{badhair}. 
\begin{figure}[H]
    \centering
  \begin{tikzpicture}[scale=1.2, every node/.style={font=\small}]
    \coordinate (z) at (0,0);
    \coordinate (y) at (2,0);
    \coordinate (x) at (1,1.5);
    \coordinate (u) at (3,0);
    \coordinate (w) at (4,0);

    \draw[gray, thick] (z) -- (y) -- (x) -- (z);
    \draw[gray, thick] (y) -- (u) -- (w);

    \foreach \p in {x, y, z, u, w}
        \fill[blue!80!black] (\p) circle (2pt);

    \node[above] at (x) {$E_\zeta(x)=4$};
    \node[below] at (y) {$E_\zeta(y)=3$};
    \node[below] at (z) {$E_\zeta(z)=2$};
    \node[above] at (u) {$E_\zeta(u)=U$};
    \node[below] at (w) {$E_\zeta(w)=5$};

    \draw[->, thick, blue!60!black]
        ([shift={(-0.95,1)}]y) arc[start angle=90, end angle=210, radius=0.5];

    \node at (0.8,0.6) {$\cal E$};

\end{tikzpicture}
    \caption{\small{Graph (with state space $K=\{x,y,z,u,w\}$) consisting of a triangle connected to a line.}}
    \label{badhair}
\end{figure}
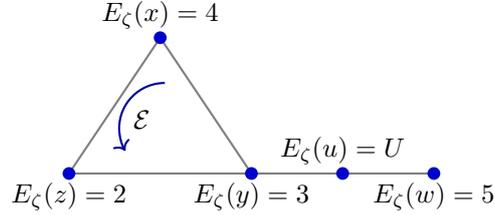
The transition rates are  
\begin{align*}
   k_\lambda(n,n')&= \frac{\upsilon}{1+e^{-\beta(E_\zeta(n)-E_\zeta(n')+s_\lambda(n,n'))}}, \qquad n,n' \in \{w,u,x,y,z \} \\
     s_\lambda(x,z)&=s_\lambda(z,y)=s_\lambda(y,x)=\cal E\quad \quad s_\lambda(u,w)=s_\lambda(y,u)=0
\end{align*}
 and the edge reactivities can be computed from \eqref{a = sqrt k}. 

In what follows, we set the frequency $\upsilon=1$ as a time reference
for all jumps. The energy levels $E_\zeta(n)$ are indicated in Fig.~\ref{badhair}, and we keep $E_\zeta(u) = U$ as a free parameter. We use arbitrary units, which involve the temperature scale but are of no relevance here.

Taking control parameters $\lambda=(\beta, U, \cal E)$, we solve the Poisson equation \eqref{Pois} for the source \eqref{activity} to obtain quasipotentials for the excess dynamical activity \eqref{VA}. For example,
\[
A_\lambda(x)= \sqrt{k(x,y)\cdot k(y,x)}+\sqrt{k_\lambda(x,z)\cdot k_\lambda(z,x)}=\frac{e^{\beta  (\varepsilon/2 +1)}}{\big(e^{\beta  \varepsilon }+e^{2 \beta }\big)}+\frac{1}{2} \text{sech}\left(\frac{1}{2} \beta  (\varepsilon +1)\right).
\]
The corresponding quasipotentials  are obtained from $L_\lambda V_\lambda^\cal A=\langle A_\lambda\rangle_\lambda^{\mathrm{s}}-A_\lambda $, and they are shown in Fig.~\ref{vactivity58}.\\ 
Note that at vanishing temperature, the quasipotential $V^A_\lambda$ remains bounded for $U=5$, while $V^A_\lambda(w)$ and $V^A_\lambda(u)$ diverge for $U=8$.
\begin{figure}[ht]    
\centering
      \begin{subfigure}{0.49\textwidth}  
		\includegraphics[scale=0.8]{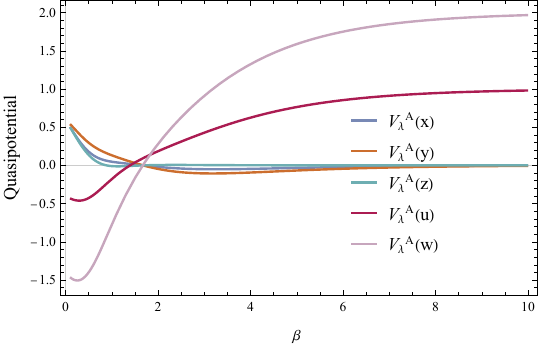} 
\caption{{ $U=5$. }}
	\end{subfigure}
  \centering
      \begin{subfigure}{0.49\textwidth}
		\includegraphics[scale=0.82]{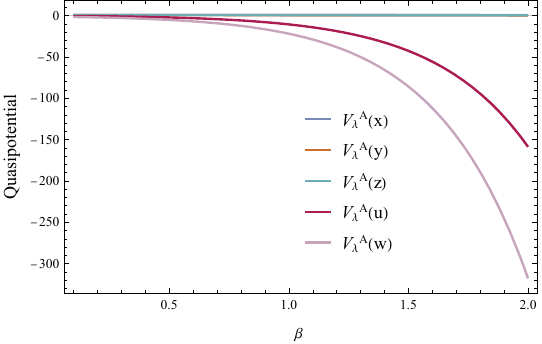}  
  \caption{{ $U=8$. }}
	\end{subfigure}
\caption{\small{Quasipotential $V^A_\lambda$ for all states $\{x,y,z,u,w\}$ as a function of inverse temperature for the model in \fig\ref{badhair} with $\cal E=1$ for (a) $U=5$ and (b) $U=8$. Increasing the energy $U$ causes localization at the outer edge of the graph, which leads to the low-temperature divergence of $V^A_\lambda(u)$ and $V^A_\lambda(w)$. The differences between $V^A_\lambda(x), V^A_\lambda(y)$ and $V^A_\lambda(z)$ are invisible on  that scale.}}\label{vactivity58}
\end{figure}
The reason has to do with the increasingly long relaxation time for large $U$: the dominant low-temperature state for $\cal E\leq 1$ and $U>2$ is $z$ and is difficult to reach from $w$.   We discuss the more general scenario in Section \ref{section extended third law}.
\begin{figure}[H]
    \centering
    \begin{subfigure}{0.49\textwidth}
        \centering
        \includegraphics[scale=0.82]{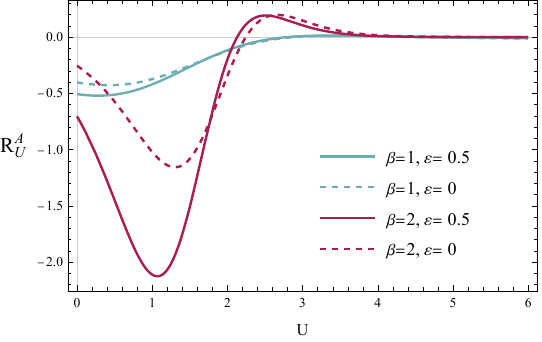}
        \caption{\small{$\cal E$=0.5 and $\cal E$=0}}
    \end{subfigure}
    \hfill
    \begin{subfigure}{0.49\textwidth}
        \centering
        \includegraphics[scale=0.82]{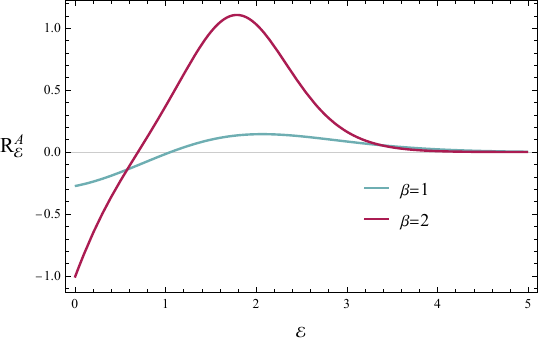}
        \caption{\small{$ U=8$}}
    \end{subfigure}
    \caption{\small{Response coefficients $R^A_U$ and $R^A_\cal E$ for the jump process given in \fig \ref{badhair}, by varying $U$ and $\cal E$ for different temperatures.  Note how the low-temperature response, $\beta=2$, jumps up from zero at the $U \simeq 2$ value where the structure of dominant states changes. }} \label{responsesA}
\end{figure}
 When $U<2$, smaller than the energy of the other states, and $\cal E < 1$, the dominant state becomes state $u$, which influences the system’s response as evident in Fig~\ref{responsesA}(a) showing the response $R^A_U$ of the reactivity to changes in $U$. We included the equilibrium case $\cal E=0$ in Fig. \ref{responsesA}(a) showing that reactivities and dynamical activities obviously remain present in equilibrium and that real excesses after quasistatic perturbations can be treated in the same general way.\\
 Fig~\ref{responsesA}(b) shows that the driving  $\mathcal{E}$ does not significantly affect the excess reactivity for large driving, since the corresponding response $R^A_\cal E$ becomes small. 
\end{example}

\begin{example}[Excess current]\label{excessCurrent}
To illustrate how to calculate the excess current \eqref{jint}, consider the jump process on the graph in Fig.~\ref{twotriangles}, consisting of two triangles connected by bridges with transition rate~$\alpha$.  On the left triangle, there is a counter-clockwise driving~$\mathcal{E}>0$. 

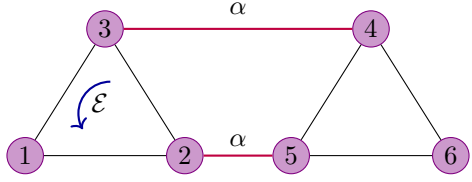
\begin{figure}[H]
\centering
       \begin{tikzpicture}[scale=1.4, every node/.style={circle, draw=Purple, fill=Purple!40, inner sep=2pt}]
    \node (1) at (0,2)  {1};
    \node (2) at (1.5,2) {2};
    \node (3) at (0.75,3.2) {3};
    \node (4) at (3.25,3.2) {4};
    \node (5) at (2.5,2) {5};
    \node (6) at (4,2) {6};

    \draw (1) -- (2);
    \draw (2) -- (3);
    \draw (3) -- (1);

    \draw[thick,purple] (3) -- (4);
    \draw (4) -- (5);
    \draw (5) -- (6);
    \draw (6) -- (4);
    \draw[thick,purple]  (5) -- (2);

     \draw[->, thick, blue!60!black] 
        ([shift={(-0.7,0.7)}]2) arc[start angle=90, end angle=210, radius=0.3];

\node[draw=none, fill=none, inner sep=0pt] at (0.7,2.5) {$\mathcal{E}$};
\node[draw=none, fill=none, inner sep=0pt] at (2,2.15) {$\alpha$};
\node[draw=none, fill=none, inner sep=0pt] at (2,3.4) {$\alpha$};
\end{tikzpicture}
\captionsetup{justification=centering}
    \caption{\small{ Graph of Example \ref{excessCurrent} with left and right triangles.}}\label{twotriangles}
\end{figure}
The transition rates over the ``bridges'' are $$k_\lambda(3,4) = k_\lambda(4,3) = k_\lambda(2,5) = k_\lambda(5,2) =\alpha>0$$ while for the edges making the two triangles, we put
\begin{equation}\label{as}
k_{\lambda}(x,y) = a \,e^{\beta/2 (E(x) - E(y) + w(x,y))},
\end{equation}
where
\begin{align}\label{enw}
E_\zeta(1)&= 1, E_\zeta(2)=2, E_\zeta(3)=3,\; E_\zeta(4)= 3, E_\zeta(5) =2, E_\zeta(6) = 1, \\
w(3,1) &= w(1,2) = w(2,3) = \mathcal{E} = -w(1,3) = -w(2,1) = -w(3,2) > 0
\end{align}
and fixing $a$ gives a reference frequency to the time-scale of the currents in the loops. Even though there is no driving in the right triangle, there is a nonzero stationary ({\it i.e.}, steady) current $j_\lambda(4,6)= \rho_\la^{\mathrm{s}}(4)k_\la(4,6)- \rho_\la^{\mathrm{s}}(6)k_\la(6,4)$ there as well, \cite{affinecurrent}; see \fig\ref{reexcu}(a). 
Following the same arguments as before, when the driving $\cal E$ is slowly changed from $0 \rightarrow 1$  in the left triangle, the excess (or geometric) current in the right triangle equals 
\begin{align}\label{excu464}
        \mathcal{J}^{\text{exc}}(4,6)
   &=\int_{0}^{1} \id \cal E \   \big[k_\cal \lambda(4,6) \  \cal V_{\lambda}  (4) \ -  k_\lambda(6,4) \ \cal V_{\lambda} (6)\big] 
\end{align}
with the quasipotentials $\cal V_\lambda$ from the forward Poisson equation \eqref{poiscu1}. The result (for $a=1$) is plotted in Fig.~\ref{reexcu}.
\begin{figure}[H]
    \centering
    \begin{subfigure}{0.49\textwidth}
        \centering
        \includegraphics[scale=0.75]{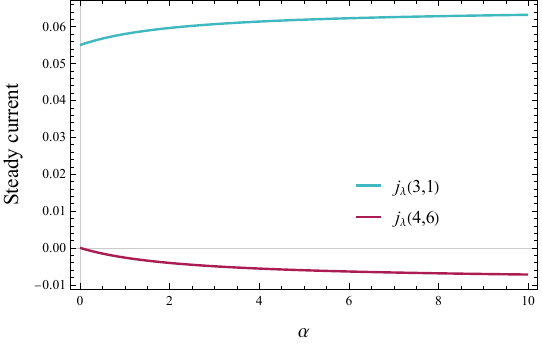}
        \caption{\small{}}
    \end{subfigure}
    \hfill
    \begin{subfigure}{0.49\textwidth}
        \centering
        \includegraphics[scale=0.75]{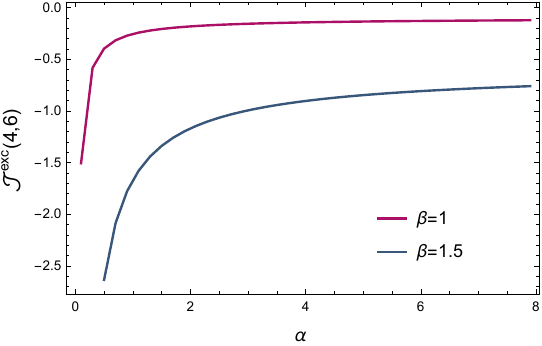}
        \caption{\small{}}
    \end{subfigure}
    \caption{\small{(a) Steady currents $j_\lambda$ (in units of $a$) in the left and right triangles in \fig \ref{twotriangles} for $\beta = 1$ and $\cal E=0.5$.  (b) Excess current $\cal J^{\text{exc}}$ (also in units of $a$) in the right triangle as a function of $\alpha$ for $\beta=1$ and $\beta=1.5$, where $\cal E$ varies from  $0 $ to $ 1$.} }\label{reexcu}
\end{figure}
\end{example}

\section{Berry curvature}\label{max}
For a cyclic transformation, the curve $\Gamma = \partial \mathscr{S}$ traces the boundary of a (two-dimensional) surface $\mathscr{S}$ in the positive (counter-clockwise) orientation such that we can rewrite the excess \eqref{bpha} using Stokes' theorem, 
 \cite{DifferentialGeometryinPhysics},
\begin{align}
H^{\text{exc}}&= \oint_{\partial \mathscr{S}} \id \lambda \cdot  R(\lambda)  = \oint_{\partial \mathscr{S}} \omega  = \int_{\mathscr{S}} \id \omega = \int_{\mathscr{S}} \frac{1}{2} \Omega_{\mu \nu} \ \id \lambda^\mu \wedge  \id \lambda^\nu \label{Berry curvature},   \\
\Omega_{\mu \nu} &= \partial_\mu R_\nu - \partial_\nu R_\mu, \nonumber
\end{align}

where $ \id \lambda^\mu \wedge \id \lambda^\nu$ represents the area element on the surface $\mathscr{S}$ with wedge product $\wedge$ and we have applied the exterior derivative ``$\id $'' to obtain an antisymmetric second-rank tensor $\Omega$  ($\Omega_{\mu \nu} = -\Omega_{\nu \mu}$), called Berry curvature. The Berry potential (called Berry connection in the context of differential geometry) is the response $R(\lambda)$, and it acts as a vector potential with $\Omega$ as the corresponding magnetic field (or field strength tensor).  That analogy is used in Section \ref{section ab effect} for an Aharonov-Bohm-type effect.  \\

Using the components \eqref{A components}, the Berry curvature gets expressed as
\begin{align}
    \Omega_{\mu \nu} 
    & =-\sum_{x \in K} \left[ \partial_\mu\left(   V_\lambda(x) \ \partial_\nu  \rho_{\lambda}^{\mathrm{s}}(x) \right) - \partial_\nu \left( V_\lambda(x) \ \partial_\mu  \rho_{\lambda}^{\mathrm{s}}(x) \right) \right] \nonumber\\
    & =- \sum_{x \in K} \left[ \partial_\mu V_\lambda(x) \partial_\nu  \rho_{\lambda}^{\mathrm{s}}(x) - \partial_\nu V_\lambda(x) \partial_\mu  \rho_{\lambda}^{\mathrm{s}}(x) \right]\nonumber \\
    &= \partial_\mu \left \langle \partial_\nu V_\lambda \right \rangle_\lambda^{\mathrm{s}} - \partial_\nu \left \langle \partial_\mu V_\lambda \right \rangle_\lambda^{\mathrm{s}}. \label{berry curvature maxwell eq}
\end{align}
\begin{example_contd}[Berry curvature of the excess heat] 
    For $\mu, \nu \in \{\beta_1, \beta_2, \alpha\}$, the Berry curvature is 
    \begin{align*}
        \Omega^{\cal P_i}_{\mu \nu} = \partial_{\mu} R^{\cal P_i}_{\nu} - \partial_{\mu}  R^{\cal P_i}_{\nu}.
    \end{align*}
Considering the excess heat to the bath $i=1$,
    \begin{align}
      \Omega^{\cal P_1}_{\beta_1 \beta_2}&= -\frac{1}{\cal Z^3} \left[2 (\alpha -1) \alpha  \delta _1^2 \delta _2 e^{\beta _1 \delta _1+\beta _2 \delta _2} \left(e^{\beta _2 \delta _2} \left(2 (\alpha +1) e^{\beta _1 \delta _1}+1\right)+e^{\beta _1 \delta _1}\right) \right], \label{example 1 berry}\\
\Omega^{\cal P_1}_{\beta_1 \alpha}&=-\frac{1}{\cal Z^3} \left[\delta _1^2 e^{\beta _1 \delta _1} \left(e^{2 \beta _2 \delta _2} \left(-\alpha +4 e^{\beta _1 \delta _1}+1\right)+(\alpha +3) e^{\beta _1 \delta _1+\beta _2 \delta _2}+e^{\beta _1 \delta _1}\right) \right], \nonumber \\
\Omega^{\cal P_1}_{ \beta_2 \alpha}&= -\frac{1}{\cal Z^3} \left[\delta _1 \delta _2 e^{\beta _2 \delta _2} \left((\alpha +3) \left(-e^{\beta _1 \delta _1+\beta _2 \delta _2}\right)+e^{2 \beta _1 \delta _1} \left(\alpha -4 e^{\beta _2 \delta _2}-1\right)-e^{\beta _2 \delta _2}\right) \right]. \nonumber
    \end{align}
\end{example_contd}

\begin{example_contd}[ Berry curvature of excess reactivity] 
We consider the Berry curvature
\[
\Omega^A_{\mathcal{E} U} = \partial_{\mathcal{E}} R^A_U - \partial_U R^A_{\mathcal{E}},
\]
whose behavior is shown in \fig\ref{beryycurveBadhair} as a function of $\beta$ and driving $\cal E$.

\begin{figure}[H]
    \centering
    \begin{subfigure}{0.49\textwidth}
        \centering
        \includegraphics[scale=0.85]{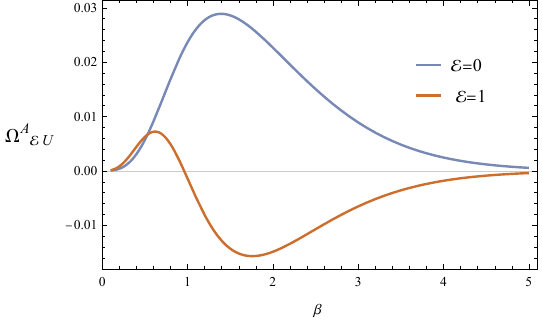}
        \caption{\small{}}
    \end{subfigure}
    \hfill
    \begin{subfigure}{0.49\textwidth}
        \centering
        \includegraphics[scale=0.85]{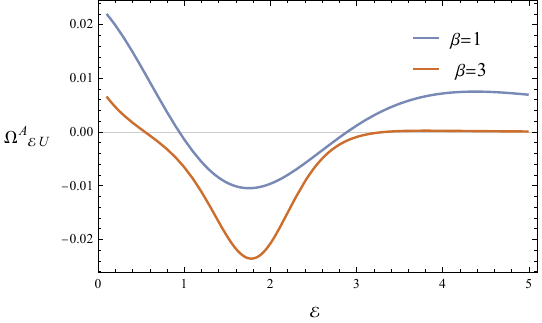}
        \caption{\small{}}
    \end{subfigure}
    \caption{\small{ Berry curvature $\Omega^A_{\mathcal{E} U}$ for the excess reactivity in the dynamics of \fig\ref{badhair} for  $U=5$. (a) as a function of the inverse temperature for $\cal E = 0$ (equilibrium)  and $\cal E =1$. Observe that the Berry curvature vanishes at low temperatures.  (b) as a function of $\cal E$ for $\beta=1$ and $\beta=3$.
    } }\label{beryycurveBadhair}
\end{figure}

\end{example_contd}

\subsection{As violation of the Maxwell relations}\label{section noneq response theory}
From \eqref{berry curvature maxwell eq} the Berry curvature is an antisymmetrization of a second-order response, and therefore, it measures the rotational part of the response. To be specific, we take the context of elasticity.
For \eqref{h state funtion} with function $h_\zeta(x,y) = E_\zeta(x) - E_\zeta(y)$, we have $f_\lambda(x) = - L_\lambda E_\zeta(x) $ and  the Poisson equation \eqref{Pois} has the simple solution $V_\lambda = E_\zeta(x) - \langle E_\zeta(x)\rangle_\lambda^{\mathrm{s}}$, which is the centered energy. The Berry potential then becomes
\begin{align}\label{spec}
    R_\mu^{E} &= \left \langle \partial_\mu V_\lambda \right \rangle_\lambda^{\mathrm{s}}  =   \left \langle \partial_\mu E_\zeta \right \rangle_\lambda^{\mathrm{s}} -  \partial_\mu \left \langle E_\zeta \right \rangle_\lambda^{\mathrm{s}} = - \left \langle E_\zeta \  \partial_\mu \log \rho_\lambda^{\mathrm{s}} \right \rangle_\lambda^{\mathrm{s}},
\end{align}
which is interpreted (up to a minus sign) as a mean force since it is the stationary expectation of the gradient of a (mean-zero) energy $E_\zeta(x) - \langle E_\zeta(x)\rangle_\lambda^{\mathrm{s}}$.  The corresponding stiffness matrix is\footnote{The definition of the stiffness matrix follows \cite{Maes_2019, Demaerel_2018} where it describes the linear stability of a probe $x$, coupled to a bath $y$ through a potential $U(x,y)$, around a fixed point $x^*$ where the induced force vanishes $\vec{f}(x^*) = - \langle \partial_{\vec{x}} U(x,y) \rangle_x = 0$. Here $\langle \cdot \rangle_x$ denotes the expectation value over the (possibly nonequilibrium) medium with the probe fixed at $x(t) = x$. The linear response to a perturbation is then governed by
    \begin{align*}
        f_\nu(x) &= \partial_{x^\mu} f_\nu(x^*) (x^\mu - x^{*\mu}) + O((x-x^*)^2) \\
        &= - \partial_{x^\mu} \left \langle \partial_{x^\nu} U(x,y) \right \rangle_x(x^\mu - x^{*\mu}) + O((x-x^*)^2), 
    \end{align*}
    where one recognises the induced stiffness matrix (in the sense of Hooke's Law)
    \begin{align*}
        \cal K_{\mu \nu} = \partial_{x^\mu} \left \langle \partial_{x^\nu} U(x,y) \right \rangle_x.
    \end{align*}
In equilibrium, it is symmetric due to the Maxwell symmetry relations, but out of equilibrium, the matrix  $\cal K$ generally picks up an antisymmetric part, representing rotational forces. By analogy, we also call $\partial_\mu \left \langle \partial_\nu V_\lambda \right \rangle_\lambda^{\mathrm{s}}$ the stiffness matrix corresponding to the quasipotential $V_\lambda$. }
\begin{align*}
    \cal K_{\mu \nu} = \partial_{\mu} R_\nu^E = \partial_\mu \left \langle \partial_\nu V_\lambda^E \right \rangle_\lambda^{\mathrm{s}} = \langle \partial_{\mu} \partial_\nu E_\zeta\rangle_\lambda^{\mathrm{s}} - \partial_{\mu} \partial_\nu \left \langle E_\zeta \right \rangle_\lambda^{\mathrm{s}} + \langle \partial_\mu \log \rho_\lambda^{\mathrm{s}} \ \partial_\nu E_\zeta \rangle_\lambda^{\mathrm{s}}.
\end{align*}
A sufficient condition for local stability is the positivity of $\cal K$, and that depends only on its symmetric part, \cite{Maes_2019, Demaerel_2018}. Its antisymmetric part is proportional to the Berry curvature \eqref{berry curvature maxwell eq},
\begin{equation}\label{mu}
    \Omega_{\mu \nu}^E = \partial_\mu \left \langle \partial_\nu V_\lambda^E \right \rangle_\lambda^{\mathrm{s}} - \partial_\nu \left \langle \partial_\mu V_\lambda^E \right \rangle_\lambda^{\mathrm{s}} = \mathcal{K}_{\mu \nu} - \mathcal{K}_{\nu \mu}. 
\end{equation}
Therefore, the Berry curvature measures the rotational part of the mean force. More generally, for other choices of flux $f_\lambda$, the $\cal K$  measures the dependence of the response on another parameter. In that sense, a nonzero Berry curvature signifies a breaking of the Maxwell thermodynamic relations, \cite{Landaustatmech, pippard1966elements}.  Indeed, the latter follow under equilibrium conditions from the equality of mixed second derivatives, independent of their order, of thermodynamic potentials, while $\Omega$ in \eqref{berry curvature maxwell eq} measures the failure for $\partial_\mu$ and $\partial_\nu$ to commute.\\

Expression \eqref{mu} makes the Berry curvature an antisymmetrization of a response function. To get more explicit expressions, we need to compute how (the in general unknown) $\rho_\lambda^{\mathrm{s}}$ changes with respect to the parameters $\lambda = (\beta, \zeta)$: 
 \begin{equation}\label{su}
    \mathcal{K}_{\mu \nu} = \partial_\mu \left \langle \partial_\nu V_\lambda \right \rangle_\lambda^{\mathrm{s}} = \langle \partial_{\mu} \partial_\nu  V_\lambda\rangle_\lambda^{\mathrm{s}} + \langle \partial_\mu \log \rho_\lambda^{\mathrm{s}} \ \partial_\nu V_\lambda\rangle_\lambda^{\mathrm{s}}.
\end{equation}   
Since the first term is symmetric, it is the last term that is of interest for the Berry curvature. It can be computed in a variety of ways, also depending on how the $\lambda$ enters the dynamics. We continue with the choice \eqref{h state funtion} as in \eqref{spec} where $V_\lambda = E_\zeta(x) - \langle E_\zeta(x)\rangle_\lambda^{\mathrm{s}}$.  We assume that the parameters $\zeta$ that are varied only enter the energy $E_\zeta$ in the sense that, for \eqref{dec}, $s_\lambda(x,y) = \beta[E_\zeta(x)- E_\zeta(y) + w(x,y)]$ for some antisymmetric $w(x,y)=-w(y,x)$ and
\begin{equation}\label{unp}
k_\lambda(x,y) = a(x,y) \,e^{\beta\,[E_\zeta(x)- E_\zeta(y) + w(x,y) ] /2 }.
\end{equation}
In other words, the parameters $\zeta$  do not directly interfere with the driving $w(x,y)$, nor with the reactivities $a(x,y)$.  We then start the usual treatment of linear response around nonequilibrium \cite{Baiesi2009, Baiesi2009_2, Baiesi_2013, pei2024inducedfrictionprobemoving,frenesy, Maes_2020} by writing
\begin{align}
\langle \partial_\mu \log \rho_\lambda^{\mathrm{s}} \ \partial_\nu V_\lambda\rangle_\lambda^{\mathrm{s}} & = \lim_{\ell \downarrow 0} \frac 1{\ell}\,\left[\ \sum_{x \in K} \partial_\nu V_\lambda(x) \ \rho_{\lambda+ \ell}^{\mathrm{s}}(x) - \sum_{x \in K} \partial_\nu V_\lambda(x) \ \rho_{\lambda}^{\mathrm{s}}(x)\right]  \notag   \\
& = \lim_{\ell \downarrow 0}  \lim_{t\downarrow 0}\frac 1{\ell}\,\left[\ \sum_{x \in K} e^{tL_\lambda} \partial_\nu V_\lambda(x) \ \rho_{\lambda+ \ell}^{\mathrm{s}}(x) - \sum_{x \in K} e^{tL_\lambda} \partial_\nu V_\lambda(x) \ \rho_{\lambda}^{\mathrm{s}}(x)\right],   \label{2te} 
\end{align}
where $\ell = \id \zeta^\mu$ changes $\zeta$ in the $\mu$-parameter direction only. Moreover, we have inserted the generator $L_\lambda$ 
because the first term in \eqref{2te} can be written as
\[
\sum_{x \in K} e^{tL_\lambda} \partial_\nu V_\lambda(x) \  \rho_{\lambda+\ell}^{\mathrm{s}}(x) =\left \langle \partial_\nu V_\lambda(x(t))\right \rangle^*,
\]
where the $*-$process is a time-dependent perturbation of the steady nonequilibrium process corresponding to the second term in \eqref{2te} with rates \eqref{unp} at fixed $\zeta$. More precisely, the average $\langle \cdot  \rangle^*$ defined as
\begin{itemize}
    \item starts far in the past ($t=-\infty$), with $x$ distributed by $\rho_\lambda^{\mathrm{s}}$;
    \item for $t<0$, the dynamics runs with parameter values $\zeta + \ell$ till time $t=0$;
    \item switches back to the dynamics with \eqref{unp} at fixed $\zeta$ for $t\geq 0$.  
\end{itemize}
A standard calculation, using techniques from \cite{Baiesi2009, Baiesi2009_2, Baiesi_2013, pei2024inducedfrictionprobemoving,frenesy, Maes_2020}, then gives
\begin{align}
 \langle \partial_\mu \log \rho_\lambda^{\mathrm{s}} \ \partial_\nu V_\lambda\rangle_\lambda^{\mathrm{s}} & = 
     \frac{\beta}{2} \left[\left \langle \partial_\mu E_\zeta \  ; \ \partial_\nu V_\lambda \right \rangle_\lambda^{\mathrm{s}} - \int_{- \infty}^0 \id \tau \  \left \langle L_\lambda\partial_\mu E_\zeta(\tau) \ ; \ \partial_\nu V_\lambda(0) \right \rangle_\lambda^{\mathrm{s}}\right]\label{stt}
\end{align}
Here $\langle \cdot \ ; \ \cdot \rangle_\lambda^{\mathrm{s}}$ indicates the covariance in the stationary distribution, given by
\begin{align*}
    \langle f \ ; \ g \rangle_\lambda^{\mathrm{s}} &= \langle f \cdot g \rangle_\lambda^{\mathrm{s}} - \langle f \rangle_\lambda^{\mathrm{s}} \cdot \langle  g \rangle_\lambda^{\mathrm{s}} \\
    & = \sum_{x \in K} f(x) \ g(x) \  \rho_\lambda^{\mathrm{s}}(x) - \left(\sum_{x \in K} f(x) \ \rho_\lambda^{\mathrm{s}}(x) \right) \cdot \left(\sum_{x \in K} g(x) \ \rho_\lambda^{\mathrm{s}}(x) \right)
\end{align*}

From a pathspace analysis, the first term in the right-hand side of \eqref{stt} represents the entropic contribution, while the second term in \eqref{stt} is frenetic, \cite{Maes_2020}. Hence, the Berry curvature decomposes as well into
\begin{align}
    \Omega_{\mu \nu} &= \langle \partial_\mu \log \rho_\lambda^{\mathrm{s}} \ \partial_\nu V_\lambda\rangle_\lambda^{\mathrm{s}} - \langle \partial_\nu \log \rho_\lambda^{\mathrm{s}} \ \partial_\mu V_\lambda\rangle_\lambda^{\mathrm{s}} = \Omega_{\mu \nu}^{\text{ent}} + \Omega_{\mu \nu}^{\text{fren}}, \notag\\
   \Omega_{\mu \nu}^{\text{ent}}&=  \frac{\beta}{2} \left[ \left \langle \partial_\mu E_\zeta \  ; \ \partial_\nu V_\lambda  \right \rangle_\lambda^{\mathrm{s}} -  \left \langle \partial_\nu E_\zeta \  ; \ \partial_\mu V_\lambda  \right \rangle_\lambda^{\mathrm{s}}  \right], \label{ento}\\
   \Omega_{\mu \nu}^{\text{fren}} & = -\frac{\beta}{2} \int_{- \infty}^0 \id \tau \ \left[  \left \langle L_\lambda\partial_\mu E_\zeta(\tau) \ ; \  \partial_\nu V_\lambda(0) \right  \rangle_\lambda^{\mathrm{s}} - \left \langle L_\lambda\partial_\nu E_\zeta(\tau) \ ; \  \partial_\mu V_\lambda(0) \right  \rangle_\lambda^{\mathrm{s}} \right]\notag
\end{align}
and it is easily checked that the entropic part $ \Omega_{\mu \nu}^{\text{ent}} =0$ vanishes for $V_\lambda = E_\zeta(x) - \langle E_\zeta(x)\rangle_\lambda^{\mathrm{s}}$.  In other words, it is the frenetic contribution that is responsible for the violation of the Maxwell relations.  In that same spirit, in the next subsection \ref{section close to equilibrium}, we show that the vanishing of the  Berry curvature for excess entropy is equivalent to the Clausius heat theorem, \cite{clausius1865,kom, Maes_2013, closeeq, Komatsu_2008, Bertini_2012, Bertini_2013}.

\subsection{Equilibrium case}\label{section close to equilibrium}
We recall from \eqref{rent} the excess entropy change
\begin{align}
   \cal  S_\ve^{\text{exc}}& = \int_0^{\tau/\ve} \id t \sum_{x \in K} \Sigma_{\lambda(\ve t)}(x) \left(\rho_t^\ve(x) - \rho^{\mathrm{s}}_{\lambda(\ve t)} (x)\right)  \notag\\ 
   \cal  S^{\text{exc}} &= \lim_{\ve \downarrow 0} \cal  S_\ve^{\text{exc}}  = 
    \int_{\Gamma} \left \langle \nabla_\lambda V^{\Sigma}_\lambda \right \rangle_\lambda^{\mathrm{s}} \cdot \id \lambda  =- \int_\Gamma \id  \lambda\cdot \left \langle V^{\Sigma}_{\lambda}
    \, \nabla_\lambda \log \rho^{\mathrm{s}}_\lambda \right \rangle_\lambda^{\mathrm{s}},\label{limit excess entropy}
\end{align}
where the quasipotential $V^{\Sigma}_\lambda$ solves
\begin{align}\label{Vsigma}
   V^{\Sigma}_\lambda(x) = \int_0^\infty \id t \ e^{t L_\lambda} \left[\Sigma_\lambda(x) - \left \langle  \Sigma_\lambda\right \rangle_\lambda^{\mathrm{s}} \right].
\end{align}
Under (global) detailed balance, there exists an energy function $E_\zeta$ with corresponding equilibrium free energy $\cal F_\lambda$ and $\rho_\lambda^{\mathrm{s}} \propto e^{- \beta E_\zeta}$. The heat sent to the heat bath during the transition $x\rightarrow y$, and the associated quasipotential are given by
\[ q_\lambda(x,y) = k_B T s_\lambda(x,y) = E_\zeta(x) - E_\zeta(y) , \qquad  V_\lambda^{\Sigma, \text{eq}}(x)=  \beta \,\left(E_\zeta(x) - \langle  E_\zeta\rangle_\lambda^{\text{eq}}\right) 
\]
such that  the response (or Berry potential) \eqref{ala} is equal to
\begin{align}\label{nabla S}
    \left \langle \nabla_\lambda V_\lambda^{\Sigma, \text{eq}}  \right \rangle_\lambda^{\text{eq}} = - \nabla_\lambda \left(\beta^2 \partial_\beta \cal F_\lambda \right) = - \nabla_\lambda \big(S_\lambda/k_B\big),
\end{align}
where $S_\lambda$ is the equilibrium entropy. 
Therefore, the response coefficient $R^{\Sigma}(\lambda) =\left \langle \nabla_\lambda V_\lambda^{\Sigma, \text{eq}}  \right \rangle_\lambda^{\text{eq}}$ is a gradient, and the Berry curvatures vanish  $\Omega_{\mu \nu}^\Sigma = 0$; hence both the  entropic and the frenetic part in \eqref{ento} vanish separately. 
That result also follows by noting that the equilibrium stiffness matrix,
\begin{align}
    \mathcal{K}_{\mu \nu}^{\text{eq}} &= - \partial_{\mu} \partial_{\nu}\left(\beta^2 \partial_\beta \mathcal{F}_\lambda \right) \notag\\
    &= \beta \langle \partial_{\mu\nu} E_\zeta\rangle^\text{eq}_\lambda -\beta^2 \langle \partial_\mu E_\zeta \,;\,\partial_\nu E_\zeta\rangle^\text{eq}_\lambda - \beta  \partial_{\mu \nu}  \langle E_\zeta\rangle_\lambda^{\text{eq}} + \delta_{\lambda^\nu, \  \beta} \  \delta_{\lambda^\mu, \ \beta} \  \text{Var}(E_\zeta)_\lambda^{\text{eq}} \label{recov}
\end{align}
is symmetric, in agreement with the Maxwell relations\footnote{That stiffness is different from the one in \cite{Maes_2019} due to the last term $\partial_{\mu \nu} \langle E_\zeta\rangle_\lambda^{\text{eq}}$.   Moreover, $\mathcal K_{\mu\nu}$ is closely related to the quantum metric \cite{Provost1980, quantummetric}. From \eqref{nabla S}, the equilibrium stiffness matrix reads $\mathcal K_{\mu \nu}^{\text{eq}} = - \partial_{\mu} \partial_\nu\left(S_\lambda/k_B \right)$, which is clearly symmetric. In the thermodynamic Ruppeiner geometry \cite{rupp1, rupp2}, a form of information geometry, the metric tensor describing the distance between two equilibrium
states is given by the negative Hessian of the entropy function, $g_{\mu \nu} = - \partial_{\mu} \partial_\nu S_\lambda$. Hence, up to the factor $k_B$, $g_{\mu\nu}$ coincides with $\mathcal K_{\mu\nu}^{\text{eq}}$. Out of equilibrium, however, this relation  generally no longer holds, and a corresponding metric structure is not evident.
}.  \\
Of course, in equilibrium, there is no housekeeping heat (zero mean entropy flux rate), implying $\cal S^{\text{exc, eq}} = \cal S^{\text{eq}}$ and we find
\begin{align}\label{clausius statement}
      & \cal S^{\text{eq}}  = -\oint_{\partial \mathscr{S}} \id \lambda \cdot \nabla_\lambda \left(\frac{S_\lambda}{k_B} \right) = - \frac{1}{k_B} \oint_{\partial \mathscr{S}} \id S_\lambda = 0
\end{align}
For the opposite direction, if all Berry curvatures are zero over the entire parameter space, the integral of the Berry potential over any loop vanishes, as follows from \eqref{Berry curvature}.  Hence, the Berry potential 
\begin{equation}\label{cl}
 R^\Sigma(\lambda) = \left \langle \nabla_\lambda \left[\beta \,\left(E_\zeta(x) - \langle  E_\zeta\rangle_\lambda^\text{s}\right)\right]  \right \rangle_\lambda^{\text{s}} = \beta \left \langle \nabla_\lambda \left(E_\zeta(x) - \langle  E_\zeta\rangle_\lambda^\text{s}\right)  \right \rangle_\lambda^{\text{s}}
\end{equation}
must be a gradient, $R^\Sigma(\lambda) = \nabla_\lambda \cal G_\lambda$. It means that
\[
\beta \left( \langle  \id E\rangle_\la - \id \langle  E\rangle_\la \right) = \id \cal G_\lambda
\]
must be a total differential, which is the Clausius relation.

\subsection{Aharonov-Bohm-type effect}\label{section ab effect}
One reason in general for the study of the Berry phase lies in its importance for topological effects.  The Aharonov-Bohm phenomenon is a standard application, \cite{Abeffect, Sakurai_Napolitano_2020}: the wave function of an electrically charged particle and its Berry phase are affected by the electromagnetic potentials ($\varphi$ and $\mathbf{A}$), despite being confined to a region where both electric $\mathbf{E}$ and magnetic fields $\mathbf{B}$ are zero. The closest analogy we can make here is to give an example where the system parameters are being slowly changed over a loop where the driving is zero, and the Berry curvature vanishes, and yet, the Berry phase is nonzero.  No spatial Aharonov-Bohm effect is considered, in the sense that the states are energy levels. \\ 

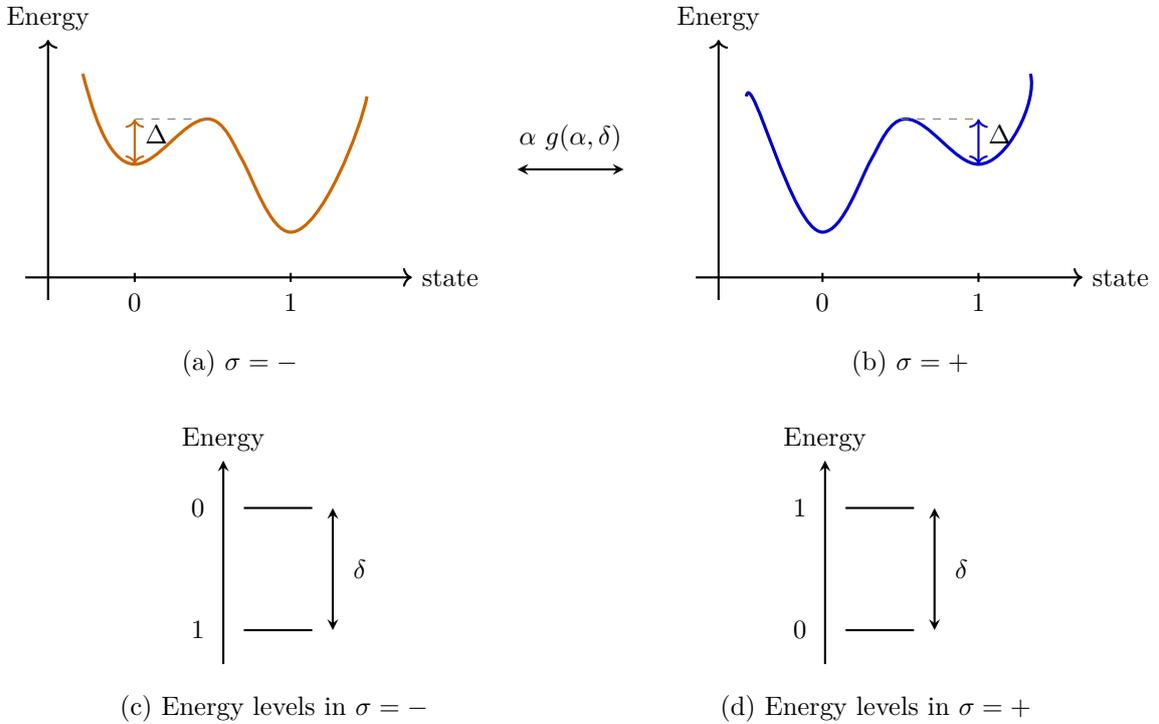
\begin{figure*}[ht]
  \centering
\subfloat[$\sigma=-$]{%
  \begin{minipage}[c]{0.4\textwidth}
    \centering
    \begin{tikzpicture}[scale=1.2]
      \def\ymax{2.5}
      \def\xmax{3.8}
      \coordinate (A) at (0.10*\xmax,0.9*\ymax);
      \coordinate (B) at (0.25*\xmax,0.5*\ymax);
      \coordinate (C) at (0.46*\xmax,0.7*\ymax);
      \coordinate (D) at (0.56*\xmax,0.52*\ymax);
      \coordinate (E) at (0.70*\xmax,0.2*\ymax);
      \coordinate (F) at (0.92*\xmax,0.8*\ymax);
      \draw[->,thick] (0,-0.1*\ymax) -- (0,1.05*\ymax) node[above] {Energy};
      \draw[->,thick] (-0.1*\ymax,0) -- (1.05*\xmax,0) node[right] {state};
      \draw[very thick,orange!80!black]
        (A) to[out=-75,in=180,looseness=0.7] (B)
        to[out=0,in=180,looseness=0.7] (C)
        to[out=0,in=120,looseness=0.8] (D)
        to[out=-60,in=180,looseness=0.6] (E)
        to[out=0,in=-100,looseness=0.5] (F);
  \draw (0.25*\xmax,0) node[below=3pt] {$0$};
\draw (0.70*\xmax,0) node[below=3pt] {$1$};
\draw[thick] (0.70*\xmax,0.05) -- (0.70*\xmax,-0.05);
    \draw[thick] (0.25*\xmax,0.05) -- (0.25*\xmax,-0.05);
     \draw[dashed,gray] (C) -- (0.25*\xmax,0.7*\ymax); 

  \draw[<->,thick,orange!80!black]
    (0.25*\xmax,0.5*\ymax) -- (0.25*\xmax,0.7*\ymax)
    node[midway,right=0.2pt,yshift = 3 pt, color = black] {$\Delta$};
    \end{tikzpicture}
  \end{minipage}
}%
\hspace{0.02\textwidth}
\begin{tikzpicture}[baseline={(0,-2)},>=stealth]
  \draw[<->,thick] (0,0) -- (1.4,0)
    node[midway,above=3pt] {$\alpha \ g(\alpha, \delta)$};
\end{tikzpicture}
\hspace{0.02\textwidth}
\subfloat[$\sigma=+$]{%
  \begin{minipage}[c]{0.4\textwidth}
    \centering
    \begin{tikzpicture}[scale=1.2]
      \def\ymax{2.5}
      \def\xmax{3.8}
      \draw[->,thick] (0,-0.1*\ymax) -- (0,1.05*\ymax) node[above] {Energy};
      \draw[->,thick] (-0.1*\ymax,0) -- (1.05*\xmax,0) node[right] {state};
      \draw[very thick,blue!80!black]
        ({\xmax - 0.10*\xmax},0.9*\ymax)
          to[out=-75,in=0,looseness=0.7]
        ({\xmax - 0.25*\xmax},0.5*\ymax)
          to[out=180,in=0,looseness=0.7]
        ({\xmax - 0.46*\xmax},0.7*\ymax)
          to[out=180,in=60,looseness=0.8]
        ({\xmax - 0.56*\xmax},0.52*\ymax)
          to[out=-120,in=0,looseness=0.6]
        ({\xmax - 0.70*\xmax},0.2*\ymax)
          to[out=180,in=80,looseness=0.5]
        ({\xmax - 0.92*\xmax},0.8*\ymax);
\draw[thick] (0.75*\xmax,0.05) -- (0.75*\xmax,-0.05);
    \draw[thick] (0.3*\xmax,0.05) -- (0.3*\xmax,-0.05);
\draw (0.3*\xmax,0) node[below=3pt] {$0$};
\draw (0.75*\xmax,0) node[below=3pt] {$1$};
      \draw[dashed,gray] (C)+(0.25,0) -- (0.75*\xmax,0.7*\ymax); 

 \draw[<->,thick,blue!80!black]
    (0.75*\xmax,0.5*\ymax) -- (0.75*\xmax,0.7*\ymax)
    node[midway,right, yshift = 3 pt, color = black] {$\Delta$};
    \end{tikzpicture}
    
  \end{minipage}
}\\[1em]
  \subfloat[\small{Energy levels in $\sigma=-$}]{%
    \begin{minipage}[c]{0.45\textwidth}
      \centering
      \begin{tikzpicture}[scale=0.9,>=stealth]
        \draw[->,thick] (0,0) -- (0,3) node[above] {Energy};
        \draw[thick] (0.3,0.5) -- (1.3,0.5);
        \draw[thick] (0.3,2.3) -- (1.3,2.3);
        \node[left=6pt] at (0.1,0.5) {$1$};
        \node[left=6pt] at (0.1,2.3) {$0$};
        \draw[<->,thick] (1.6,0.5) -- (1.6,2.3)
          node[midway,right=4pt] {$\delta$};
      \end{tikzpicture}
    \end{minipage}
  }\hspace{0.04\textwidth}%
  \subfloat[\small{Energy levels in $\sigma=+$}]{%
    \begin{minipage}[c]{0.45\textwidth}
      \centering
      \begin{tikzpicture}[scale=0.9,>=stealth]
        \draw[->,thick] (0,0) -- (0,3) node[above] {Energy};
        \draw[thick] (0.3,0.5) -- (1.3,0.5);
        \draw[thick] (0.3,2.3) -- (1.3,2.3);
        \node[left=6pt] at (0.1,0.5) {$0$};
        \node[left=6pt] at (0.1,2.3) {$1$};
        \draw[<->,thick] (1.6,0.5) -- (1.6,2.3)
          node[midway,right=4pt] {$\delta$};
      \end{tikzpicture}
    \end{minipage}
  }
  \caption{\small{Two-level switch, the ground state is separated from the excited state with an energy gap  $\delta \geq 0$, and an energy barrier $\Delta$.  The switch rate is a function $\alpha\,g(\alpha, \delta)$ of two parameters.}}
   \label{switch1 hole}
\end{figure*}

   We continue working with the excess entropy in \eqref{limit excess entropy} with quasipotential \eqref{Vsigma}. We consider a two-level system that switches between two energy profiles (indicated as $+$ and $-$ occupation levels). The colloidal setup is represented in  Fig.~\ref{switch1 hole}(a-b) with energy gap $\delta$ and an energy barrier $\Delta\geq 0$. The states are denoted by $x=(i,\sigma)$ with $i \in \{0,1\},  $ $ \sigma=\pm$, and the transition rates are 
\begin{eqnarray}
k_\lambda((i, \sigma)   ,  (1-i,\sigma))& =& e^{-\sigma\frac{\beta \delta}{2} (1-2 i) - \beta \Delta }\notag \\  k_\lambda((i,\sigma) ,  (i,- \sigma) &=&  k_\lambda((i,-\sigma)  , (i,\sigma) ) = \al \ g(\alpha, \delta),
\end{eqnarray}
where $\alpha g(\delta, \alpha),\  \alpha \geq 0$, is an effective switching rate that, when nonzero, brings the system out of equilibrium.  We take $g(\alpha, \delta) = \mathbf{1}_{\mathcal{B}}(\alpha, \delta)$ with indicator function $\mathbf{1}_{\mathcal{B}}$.
In other words, we make a ``hole'' in parameter space by allowing switching (and hence nonequilibrium contributions) only within the bounded region $\cal B$, representing inhibition of a reaction outside $\cal B$.  The simplest case corresponds to a circular hole with center $(\alpha_c, \delta_c)$ and radius $0<r \leq \min(\alpha_c, \delta_c)$, 
\begin{align*}
   g(\alpha, \delta) =  \begin{cases}
        1 \quad \text{if } (\alpha-\alpha_c)^2+(\delta-\delta_c)^2 \leq r^2 \\
        0 \quad \text{if } (\alpha-\alpha_c)^2+(\delta-\delta_c)^2 > r^2
    \end{cases}    
\end{align*}
That function can be made (more) continuous, but it does not change the main idea.

For control parameters $\lambda$, we take the energy gap $\delta \geq 0$ and the flipping rate $\alpha \geq 0$. 
The stationary distribution is
\begin{align*}
     \rho^{\mathrm{s}}_\lambda(0, -) &=  \rho^{\mathrm{s}}_\lambda(1,+) = \frac{\alpha  e^{\frac{1}{2} \beta  (2 \Delta
   +\delta )} g(\alpha ,\delta ) +1}{2 \mathscr{Z}}, \qquad   \rho^{\mathrm{s}}_\lambda(0, +) =  \rho^{\mathrm{s}}_\lambda(1,-) =  \frac{\alpha  e^{\frac{1}{2} \beta  (2 \Delta
   +\delta )} g(\alpha ,\delta ) + e^{\beta \delta }}{2 \mathscr{Z}} , 
\end{align*}
depending on the hole size and with normalization $  \mathscr{Z} = 2 \alpha  e^{\frac{1}{2} \beta  (2 \Delta
   +\delta )} g(\alpha ,\delta )+e^{\beta  \delta }+1 $.  When  $(\alpha, \delta) \not \in  \cal B$, 
   that the stationary distribution is the equilibrium one (where $\alpha=0$).
In other words, the stationary process satisfies detailed balance for $(\alpha, \delta) \not\in \cal B$, while it is out of equilibrium when $(\alpha, \delta) \in \cal B$.\\
Using \eqref{A components},  the Berry potentials  for the excess entropy flux \eqref{limit excess entropy} become
\begin{align*}
    R_\alpha^{\cal S} = &  - \frac{\beta \delta}{\mathscr{Z}^3} \left(\alpha  \partial_\alpha g(\alpha ,\delta)+g(\alpha ,\delta )\right) e^{\frac{1}{2} \beta  (2 \Delta +\delta )} \left(e^{2 \beta  \delta }-1\right)  \\
   R_\delta^{\cal S} = &  \frac{\beta  \delta }{2 \mathscr{Z}^3}  e^{\frac{\beta  \delta }{2}} \left(e^{\beta  \delta }+1\right)  \left[\alpha  e^{\beta  \Delta } \left(\beta  \left(e^{\beta 
   \delta }+1\right) g(\alpha ,\delta )-2 \left(e^{\beta  \delta }-1\right) \partial_\delta g(\alpha ,\delta)\right)+2 \beta  e^{\frac{\beta 
   \delta }{2}}\right],
\end{align*}
leading to the Berry curvature
\begin{align}\label{berry curvature ab effect}
      \Omega_{\alpha \delta}^{\cal S} =   - \Omega_{\delta \alpha }^{\cal S} =\frac{\beta}{\mathscr{Z}^3} \,  \left(\alpha 
   \partial_\alpha g(\alpha ,\delta)+g(\alpha ,\delta )\right)\, e^{\frac{1}{2} \beta  (2 \Delta +\delta )} \left(-2 \beta  \delta  e^{\beta  \delta }+e^{2 \beta  \delta }-1\right).
\end{align}
Note that $ \Omega_{\alpha \delta}^{\cal S} = 0$ for $(\alpha, \delta) \not \in \cal B$ but $ \Omega_{\alpha \delta}^{\cal S} \neq 0$ for $(\alpha, \delta) \in \cal B$ such that the one form $R_\mu^{\cal S}  \id \lambda^\mu = R_\alpha^{\cal S}  \id \alpha + R_\delta^{\cal S}  \id \delta$ does not equal an exact differential since the mixed derivatives do not agree:
\begin{align*}
   \partial_\alpha R^{\cal S}_\delta - \partial_\delta R^{\cal S} _\alpha = \Omega^{\cal S}_{\alpha \delta } \neq 0 \qquad \forall \alpha, \delta \in \cal M.
\end{align*}
Hence, the excess entropy flux rate does not need to vanish as it does in equilibrium around every closed loop; see the previous Section \ref{section close to equilibrium}. Indeed, consider a closed loop $\Gamma$ in parameter space enclosing $\cal B$ such that $g(\Gamma) = 0$ and $\rho_\lambda^{\mathrm{s}} \Big|_\Gamma = \rho_\lambda^{\text{eq}}$.  Then,  even though the system is formally satisfying detailed balance along $\Gamma$ and the Berry curvature (equivalent of the magnetic field) vanishes along this curve, $\Omega_{\alpha \delta}^{\cal S} \Big|_{\Gamma} = 0$, we still have 
\begin{align*}
    \cal S^{\text{exc}} = \oint_{\Gamma} \left( R_\alpha^{\cal S} \id \alpha + R_\delta^{\cal S} \id \delta \right) = \oint_{\Gamma} \left( 0 \ \id \alpha +  \frac{\beta^2 \delta e^{\beta \delta}}{(1 + e^{\beta \delta})^2} \id \delta \right) \neq 0.
\end{align*}
That result can be considered as an analogue of the Aharonov–Bohm effect for the excess entropy flux.  However, it remains unclear whether this attempt of introducing a Aharonov-Bohm-type
effect in out-of-equilibrium jump processes may provide a concrete protocol accessible in experiment or via simulation.  More generally, topological effects in nonequilibrium statistical mechanics appear to require the presence of physically meaningful topological parameters.

\section{Low-temperature behavior}\label{section extended third law}
One standard formulation of the Nernst postulate, also a version of the Third Law of Thermodynamics, states that the thermodynamic entropy of an equilibrium system approaches a finite value as the temperature approaches absolute zero.  It implies that the equilibrium heat capacity must go to zero.  To understand an extension of that statement to open driven systems has been the subject of previous papers \cite{jchemphys,mathnernst}.  The present section generalizes the Nernst postulate even further, where again we consider nonequilibrium systems in contact with a heat bath at temperature $T$, but now for all types of excess as considered above, and isothermal quasistatic transformations. We show below that all the (excess) responses $R_\mu$ and the Berry curvatures vanish for isothermal transformations at vanishing absolute temperature. Context and conditions are needed, similar to the equilibrium case, which we list below:

\begin{enumerate}
    \item
We restrict ourselves to Markov jump processes.  That discretization is in line with the necessary quantum (and quantization) aspects that show up for physical systems at very low temperatures.  Shortly, it will become clear how, for instance, tunneling effects contribute positively as well. \\
Moreover, the transition rates of the Markov jump process need to remain bounded as the temperature reaches absolute zero, in accord with what is obtained from a Golden Rule approximation.
\item Secondly, there is an equilibrium-like condition, that the low-temperature behavior in an open neighborhood of parameters is determined equally by {\it dominant} states $x^*$ of the system.   More specifically, we suppose that for the considered parameter value, there is a set $\cal B$ of states (called dominant) and an open neighborhood $\mathcal{D}^*$ around that value\footnote{ Importantly, $\mathcal{D}^*$ does not need to be narrow. For example, in the case of \fig \ref{badhair}, one finds that for any values of  $U > 4$ and $0 \leq \cal E \leq 1.5$, the stationary probability at low temperature satisfies $\rho^{\mathrm{s}}(z) \approx 1$. This indicates that within this entire parameter range, the system  remains in the same dominant state $x^* = z$ as the temperature approaches zero.}, so that
\begin{equation}\label{domi}
    \lim_{\beta\uparrow \infty}\rho^{\mathrm{s}}_{\la=(\beta,\zeta)}(x^*) = \frac{1}{\cal N} \quad\text{ for any } \quad x^*\in \cal B, \quad  \zeta\in \cal D^*
\end{equation}
with ${\cal N} = |\cal B|$ the number of dominant states.  In other words,  we assume an open neighborhood of parameters around the ones we are considering, where the stationary distribution starts to concentrate equally on the same set of dominant states $x^*$.\\
 Note that when there is more than one dominant state ${\cal N}\geq 2$, it is perfectly possible to have a residual or nonzero current at vanishing temperature.
 Recall here that we consider only finite systems and ignore thermodynamic limits and the possible occurrence of thermal phase transitions.
\item So far, these conditions do not deviate much from the typical equilibrium statistical mechanical setup for the Third Law (except that we do not consider limits of spatial extension). For the third condition, we introduce the mean first-passage time $\tau_\lambda(x,y)$ to reach $y \in K$ when started from $x \in K$. It is the solution of
\begin{equation}\label{taup}
     \sum_{y \in K} k_\lambda(x,y)[\tau_\lambda(y,z) - \tau_\lambda(x,z)] +1 =0, \quad x\neq z, \qquad \tau_\lambda(z,z) = 0.
 \end{equation}
Our third condition and specific to the considered nonequilibrium setup is the requirement that the differences in mean first-passage times must remain finite as $T\downarrow 0$, in the sense that
\begin{equation}\label{cond4}
|\tau_\lambda(y,z) - \tau_\lambda(x,z)| \leq c_\zeta \quad  \text{ independent of } \beta
\end{equation}
whenever $k_\lambda(x,y) \neq 0$.  In other words, there is no exploding difference in the times to reach a given state $z$ when taking two different initial states $x$ and $y$.  Avoiding low-temperature localization is indeed the key to having saturation of all excess quantities.  Heuristically, it means that the local dynamical activity should remain larger than the steady current, uniformly in $T\downarrow 0$, and tunneling\footnote{modeled as extra nonzero transitions that survive the zero-temperature limit.} obviously helps. That requirement has been explained in \cite{jchemphys,mathnernst} for thermal response 
by demanding that the ratio between relaxation time and dissipation time needs to remain bounded at vanishing temperature.  
\end{enumerate}

Given the above conditions, the extended Third Law follows from the following reasoning.\\
We have seen in \eqref{A components} that the response $R_\mu = - \sum_{x \in K}  \ V_\lambda(x) \ \partial_\mu  \rho_{\lambda}^{\mathrm{s}}(x)$ is given in terms of the quasipotential $V_\la$.  Informally, under our assumptions above for isothermal quasistatic transformations, that response equals $R_\mu \simeq \sum_{x^* \in \cal B} v_\zeta(x^*) \  \lim_{\beta \uparrow \infty} \partial_\mu  \rho_{\lambda}^{\mathrm{s}}(x^*)$ as the temperature reaches absolute zero.  Because we assume \eqref{domi}, the response $R_\mu$ vanishes whenever the quasipotential $V_\lambda$ remains bounded as the temperature lowers. Indeed, if for all $x$, $V_\lambda(x)$  saturates to a finite value $v_\zeta(x)$ as $T\downarrow 0$, then $R_\mu \rightarrow 0$ must vanish  since from \eqref{domi}, $\rho_\lambda^{\mathrm{s}}(x^*)$ becomes independent of the parameters we vary in $\partial_\mu$. This result is our extended Third Law. \\
To justify the boundedness of $V_\lambda$ at low temperatures, the condition \eqref{cond4} enters.  Indeed, its main input is the following general identity for quasipotentials,
following \cite{pois},
\begin{equation}\label{difv}
    V_\lambda(x)-V_\lambda(y)= \sum_{z \in K} \, \rho^{\mathrm{s}}_\lambda(z) \, f_\lambda(z)(\tau_\lambda(y,z) -\tau_\lambda (x,z) )
\end{equation}
with mean first-passage time $\tau_\lambda(x,y)$  from \eqref{taup}. We conclude that the quasipotential remains finite (and hence the response vanishes at zero temperature) under \eqref{cond4}, when the differences in mean first-passage times 
$|\tau_\lambda(y,z) -\tau_\lambda (x,z)|$ remain finite.  We assume here that the source $f_\lambda$ remains finite, {\it e.g.}, from the boundedness of low-temperature transition rates (first condition).\\

Starting from \eqref{limit excess H}, a more formal argument proceeds via
\begin{align}\label{zero1}
\lim_{\beta \uparrow \infty} H^{\text{exc}}
&= \lim_{\beta \uparrow \infty} 
   \int_{\Gamma} \id \lambda \cdot  \big\langle \nabla_\lambda V_\lambda \big\rangle_\lambda^{\mathrm{s}} \notag \\
&= \sum_{x \in K}\lim_{\beta \uparrow \infty} 
   \int_{\Gamma} \id \lambda \cdot  \rho_\lambda^{\mathrm{s}}(x)\, 
   \nabla_\lambda V_\lambda(x)  \notag\\
&= \frac{1}{\cal N}\sum_{x^* \in \cal B}\int_{\Gamma} \id \lambda \cdot  \nabla_\lambda v_\zeta(x^*)
   =\frac{1}{\cal N} \sum_{x^* \in \cal B}\int_{\Gamma} \mathrm{d}v_\zeta(x^*) \notag\\
&=\frac{1}{\cal N}\sum_{x^* \in \cal B}[ v_{\zeta_\text{f}}(x^*) - v_{\zeta_\text{i}}(x^*)],
\end{align}
where $\zeta_\text{i}$ and $\zeta_\text{f}$ denote the initial and final points 
along the isothermal path $\Gamma$ inside $\mathcal{D}^*$. 
Thus, for a closed path where $\zeta_\text{i} = \zeta_\text{f}$ with temperature reaching absolute zero, it follows that the Berry phase vanishes.    But even for an arbitrary path, $\lim_{\beta \uparrow \infty} H^{\text{exc}} =0$ because $0 = \langle V_\la\rangle_\la^{\mathrm{s}} \rightarrow \sum_{x^*} v_\zeta(x^*)$ when indeed the $V_\lambda\rightarrow v_\zeta$ remain uniformly bounded with $\beta\uparrow \infty$. \\

\begin{figure}[ht]
    \centering
    \begin{subfigure}{0.49\textwidth}
     \centering
        \includegraphics[scale=0.8]{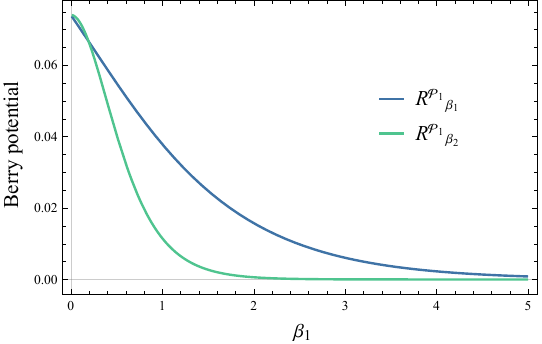}
        \caption{\small{}}
    \end{subfigure}
       \hfill
    \begin{subfigure}{0.49\textwidth}
        \centering
        \includegraphics[scale=0.8]{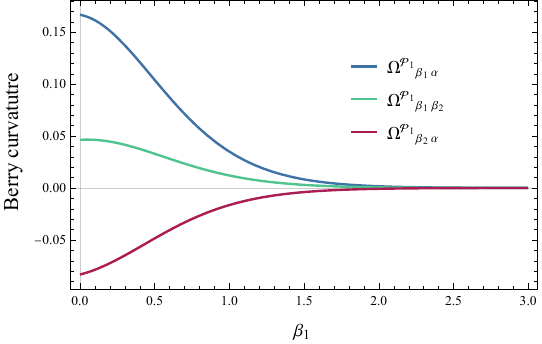}
        \caption{\small{}}
    \end{subfigure}
    \caption{\small{ Response coefficients $R^{\cal P_1}$ and Berry curvatures $\Omega^{\cal P_1}$ for the molecular conductor in \fig \ref{ziaex} with $\delta_1=1$,  $\delta_2=2$, $\beta_2=1.5\beta_1 $ and $ \alpha=1$.}}
    \label{3rexa}
\end{figure}

For the molecular conductor of Example \ref{thre}, it is seen in \fig \ref{quasiheatMC} that the quasipotentials are bounded at zero temperatures, and indeed the response coefficients and the Berry curvatures vanish at zero temperatures; see \eqref{example 1 berry} and \fig \ref{3rexa}.

\begin{figure}[ht]
    \centering
    \begin{subfigure}{0.49\textwidth}
     \centering
        \includegraphics[scale=0.8]{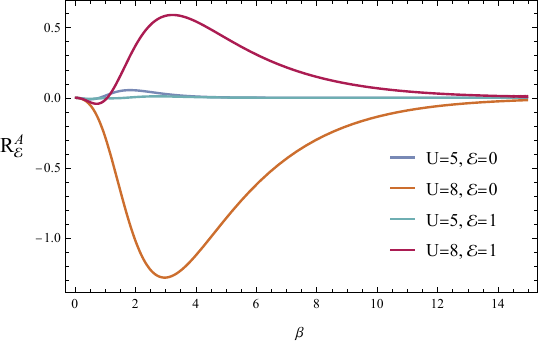}
        \caption{\small{}}
    \end{subfigure}
       \hfill
    \begin{subfigure}{0.49\textwidth}
        \centering
        \includegraphics[scale=0.8]{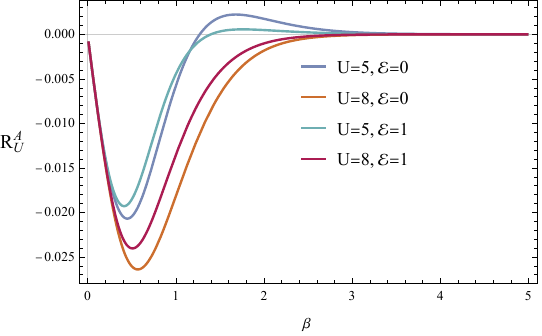}
        \caption{\small{}}
    \end{subfigure}
    \caption{\small{Response coefficients $R^A$  of  Example \ref{exre} for two different values of $U$ and $\cal E$, where $a=1$.}}
    \label{sufb}
\end{figure}

Example~\ref{exre}, corresponding to the graph in Fig.~\ref{badhair}, illustrates that the boundedness of the quasipotentials is a \emph{sufficient} but not \emph{necessary} condition. Indeed, as shown in Fig.~\ref{vactivity58}(b), the quasipotentials at zero temperature are not bounded for $U = 8$, but the response coefficients still vanish at low temperatures; see Fig.~\ref{sufb}. The reason is that here the vanishing of the stationary probabilities $\rho_\la^{\mathrm{s}}(u), \rho_\la^{\mathrm{s}}(w)$ is faster than the divergence of $V_\la^A(u), V_\la^A(w)$.
Since the response coefficients vanish at zero temperature, the Berry curvature also becomes zero, as in Fig.~\ref{beryycurveBadhair}.\\
Lastly, in the AB-type effect, one readily checks that the Berry curvature \eqref{berry curvature ab effect} vanishes for $\beta \uparrow \infty$.

It can also happen that the set $\cal B$ of dominant states changes when $\zeta$ varies. For example, we see in Fig.~\ref{responsesA}(a) that for large $\beta$ ($\beta=5$ there) the response jumps up (from being zero) at the parameter value $U\simeq 2$ where there is a change in dominant states. In that way, low-temperature response detects changes in the near-zero-temperature behavior of the system.


\section{Summary}

Excesses for quasistatic perturbations to steady nonequilibria are geometric and give rise to the analogues of Berry phase, Berry potential, and Berry curvature, quantifying the response. 
We have given various illustrations and have discussed the relation between the Berry curvature, (thermodynamic) Maxwell relations, and the (first part of the) Clausius heat theorem.  We also found a simple analogue of an Aharonov-Bohm-type effect. \\
Finally,  we have given sufficient conditions for the Berry potentials (response coefficients) and Berry curvature to vanish at absolute zero temperature, which is an extension of the Third Law.  Counterexamples reveal zero-temperature transitions.\\

\noindent 
{\bf Acknowledgments}\\
   AB is supported by the Research Foundation-Flanders (FWO) doctoral fellowship 1152725N. FK is supported by the Research Foundation--Flanders (FWO) postdoctoral fellowship 1232926N.\\ We thank Karel Netočný for initial discussions.\\
   
\textbf{Data availability statement}\\
All data that support the findings of this study are included within the article.

\bibliographystyle{unsrt}  
\bibliography{chr}
\onecolumngrid

\end{document}